\documentclass[twocolumn,superscriptaddress,showpacs,prd,aps,amsmath,amssymb,nofootinbib]{revtex4-1}
\usepackage{graphicx}
\usepackage{amssymb}
\usepackage{bm}
\usepackage{color}
\usepackage{soul}

\newcommand{\beq}{\begin{equation}} 
\newcommand{\eeq}{\end{equation}} 
\newcommand{\beqn}{\begin{eqnarray}} 
\newcommand{\eeqn}{\end{eqnarray}}

\newcommand{\zD}{{\raise1.0ex\hbox{${}^{\ \circ}$}}\!\!\!\!\!D}
\newcommand{\alone}{{\raise0.5ex\hbox{${}^{\ 1}$}}\!\!\!\!\alpha}

\newcommand{\Dl}{\Delta}

\newcommand{\nalam}{\mathrel{\raise0.9ex\hbox{$^\lambda$}\mkern-14mu
\lower0.0ex\hbox{$\nabla$}}}

\newcommand{\Nrf}{{N_r^{\rm f}}}
\newcommand{\Nrm}{{N_r^{\rm m}}}

\newcommand{\zeroD}{{\raise1.0ex\hbox{${}^{\ \circ}$}}\!\!\!\!\!D}

\newcommand{\zLap}{{\raise1.0ex\hbox{${}^{\ \circ}$}}\!\!\!\!\Delta}
\newcommand{\zna}{{\raise1.0ex\hbox{${}^{\ \circ}$}}\!\!\!\!\!\nabla}
\newcommand{\zS}{{\raise1.0ex\hbox{${}^{\ \circ}$}}\!\!\!\!\!S}

\newcommand{\carpet}{\textsc{carpet}}
\newcommand{\cocal}{\textsc{cocal}}

\newcommand{\cactus}{\textsc{cactus}}

\newcommand{\whisky}{\textsc{Whisky}}

\newcommand{\illinois}{\textsc{Illinois GRMHD}}

\newcommand{\GA}{\alpha}
\newcommand{\GB}{\beta}
\newcommand{\GG}{\gamma}
\newcommand{\GD}{\delta}
\newcommand{\GE}{\epsilon}

\newcommand{\GR}{\rho}

\newcommand{\GT}{\tau}
\newcommand{\GC}{\psi}

\newcommand{\GP}{\phi}

\newcommand{\GU}{\theta}

\newcommand{\Bphi}{\boldsymbol{\phi}}
\newcommand{\pd}{\partial}

\newcommand{\be}{\begin{equation}}
\newcommand{\ee}{\end{equation}}

\def\qeq{\mathrel{%
    \mathchoice{\QEQ}{\QEQ}{\scriptsize\QEQ}{\tiny\QEQ}%
}}
\def\QEQ{{%
    \setbox0\hbox{$I$}%
    \rlap{\hbox to \wd0{\hss--\hss}}\box0
}}

\usepackage[]{amsmath}
\usepackage[]{dsfont}
\usepackage[]{amsfonts}
\usepackage[]{amssymb}
\usepackage[]{mathrsfs}
\usepackage{times}
\usepackage{hyperref}
\hypersetup{
  colorlinks=true,        
  linkcolor=blue,         
  citecolor=cyan,         
}
\newcommand{\AJ}[3]{{Astrophys. J.} \textbf{#1}, {#2} ({#3})}

\newcommand{\arxiv}[1]{{arXiv}: {#1}}
\newcommand{\ASAS}[3]{{Astron. Astrophys.} \textbf{#1}, {#2} ({#3})}

\newcommand{\CQG}[3]{{Class. Quantum Grav.} \textbf{#1}, {#2} ({#3})}
\newcommand{\GRG}[3]{{Gen. Rel. Grav.} \textbf{#1}, {#2} ({#3})}

\newcommand{\IJMPD}[3]{{Int. J. Mod. Phys. D} \textbf{#1}, {#2} ({#3})}
\newcommand{\JCP}[3]{{J. Comput. Phys.} \textbf{#1}, {#2} ({#3})}
\newcommand{\JEM}[3]{{J. Eng. Math.} \textbf{#1}, {#2} ({#3})}

\newcommand{\MNRAS}[3]{{Mon. Not. Roy. Astron. Soc.} \textbf{#1}, {#2} ({#3})}

\newcommand{\PLA}[3]{{Phys. Lett. A} \textbf{#1}, {#2} ({#3})}

\newcommand{\PRD}[3]{{Phys. Rev. D} \textbf{#1}, {#2} ({#3})}

\newcommand{\PRL}[3]{{Phys. Rev. Lett.} \textbf{#1}, {#2} ({#3})}

\newcommand{\SIAMr}[3]{{SIAM Rev.} \textbf{#1}, {#2} ({#3})}

\begin{document}

\title{Gravitational wave content and stability of uniformly, rotating, triaxial neutron stars
in general relativity.}

\author{Antonios Tsokaros}
\affiliation{Department of Physics, University of Illinois at Urbana-Champaign, Urbana, IL 61801}

\author{Milton Ruiz}
\affiliation{Department of Physics, University of Illinois at Urbana-Champaign, Urbana, IL 61801}
\author{Vasileios Paschalidis}
\affiliation{Department of Physics, Princeton University, Princeton, NJ 08544}
\author{Stuart L. Shapiro}
\affiliation{Department of Physics, University of Illinois at Urbana-Champaign, Urbana, IL 61801}
\affiliation{Department of Astronomy \& NCSA, University of Illinois at Urbana-Champaign, Urbana, IL 61801}
\author{Luca Baiotti}
\affiliation{Graduate School of Science, Osaka University, 560-0043 Toyonaka, Japan}
\author{K\=oji Ury\=u}
\affiliation{Department of Physics, University of the Ryukyus, Senbaru, Nishihara, Okinawa 903-0213, Japan}

\date{\today}

\begin{abstract}
Targets for ground-based gravitational wave interferometers include continuous, quasiperiodic sources of 
gravitational radiation, such as  
isolated, spinning neutron stars. In this work we perform evolution simulations of uniformly rotating, triaxially 
deformed stars, the compressible analogues in general relativity of incompressible, Newtonian
Jacobi ellipsoids. We  investigate their stability and gravitational wave emission. 
We employ five models, both normal and supramassive, and track their evolution with different grid
setups and resolutions, as well as with two different evolution codes. 
We find that all models are dynamically
stable and produce a strain that is approximately one-tenth the average value of a merging binary system.
We track their secular evolution and find that all our stars evolve towards axisymmetry, maintaining
their uniform rotation, kinetic energy, and angular momentum profiles while losing their triaxiality. 
\end{abstract}

\maketitle

\section{Introduction}
\label{sec:intro}

The discovery of
gravitational waves \cite{abbott2016a} from a binary black-hole system
was a triumph that initiated a new era in astronomy and astrophysics.
Although the prime candidates for the ground-based interferometers are
binary systems, gravitational waves from isolated neutron stars also
can be detected and help reveal the nature of these objects.  Out of
the $\sim 2500$ currently known pulsars in our Galaxy, approximately
$90\%$ are isolated. Many of these single rotating stars may be
promising sources of gravitational waves \cite{EatH,
  Aasietal2013,Papaetal2016}.

A single neutron star can become an emitter of gravitational waves (GWs) as long as it has a 
nonspherical time changing quadrupole moment. 
The lack of symmetry can arise in various scenarios \cite{CA, A2010,K03}.
For example a pulsar can have a ``small mountain" that could develop following a starquake in the NS 
\cite{ST83, HJA2006}, or it can exhibit different kinds of nonspherical oscillations \cite{FS14}.
Another possibility is binary neutron star mergers, which are themselves prime candidates
for the production of gravitational radiation. When the two component stars do not have large masses 
the remnant may not undergo ``prompt'' collapse, but instead form a hypermassive star and undergo
``delayed collapse'', or form a spinning neutron star that is dynamically and secularly stable \cite{BR2016}.
At formation such remnants may be nonaxisymmetric and strong GW emitters.
A third scenario arises in gravitational stellar collapse, where the 
bouncing core can be rotating fast enough so that
nonaxisymmetric instabilities set in and deform the star into an ellipsoid \cite{LS1995}.
Fallback accretion onto newly born magnetars also supports the existence of triaxial deformations and 
the efficient production of gravitational waves \cite{PO2011}.

Despite the enormous amount of work done in the field of rotating
stars \cite{FS2013,PS2016} full general relativistic (GR) numerical
simulations that investigate the stability and accurately quantify the
GW signature of single, uniformly rotating, triaxial stars have not
been performed.  One of the reasons is the scarcity of accurate
initial models needed to study their evolution.  Typically these
objects are probed in the context of binary mergers or collapse
scenarios, which involve a substantial amount of computational
resources and make difficult a systematic parameter study. In these
cases one typically ends up with a differentially rotating object
while for single, isolated neutron stars one is often interested in
uniformly rotating stars, the GR analogues of Jacobi ellipsoids in
Newtonian theory. Such solutions have been obtained for the first time
by Nozawa in his PhD thesis \cite{N1997}, where he allowed for
azimuthal dependence in the spacetime metric, but restricted it to an
axisymmetric form. Using a different method, triaxial quasiequilibrium
models have been computed without such a restriction in the conformal
flatness approximation \cite{HMSU2008} and in the waveless
approximation \cite{UTGHSTY2016} as part of the \cocal{} code.

The ab initio computation of such nonaxisymmetric objects presents a
number of challenges. First, these objects are not stationary
equilibria, since they emit GWs, and therefore an approximate scheme
has to be applied in order to find quasistationary solutions.  This
choice has be compatible with the fact that the radiated energy within
one rotational period is much smaller than the binding energy of the
star.  Second, such models are known to exist only for stiff equations
of state.  If we assume a polytropic law $P=k\rho_0^\Gamma$, where
$\GR_0$ is the rest-mass density and $k,\ \Gamma$ are constants, then
$\Gamma$ needs to be larger than $2.24$ in the Newtonian limit
\cite{James64}. For softer equations of state mass shedding appears at
lower angular velocity than the one needed to reach the triaxial
state. GR increases the critical value of the polytropic index by a
small amount (to $\Gamma\sim 2.8$) \cite{BFG1996}.  Third, uniformly
rotating, nonaxisymmetric solutions exist only for high spin rates,
i.e. $\beta:=T/|W|$ larger than $0.14$ in the Newtonian case
\cite{Ch69}.  Here $T$ is the kinetic energy and $W$ the gravitational
potential energy.  In GR this critical value is higher \cite{SZ1996,
  BFG1998, GV2002, Saijo2006, RG2002, SL1996, YE1997}.  The
combination of the above characteristics imply that an evolutionary
follow-up will also be nontrivial, since the first challenge described
above will imply the presence of junk initial radiation, which must be
controlled, while the second and third challenges require higher
resolution than for slowly rotating stars.  Since the GW timescale to
radiate the rotational energy is $t_{\rm GW}/M \gtrsim (M/R)^{-4}$
only highly compact objects can be evolved to their endpoint state,
while lower compacteness stars can be studied only partially. High
compactness requires higher resolution, which increases the
computational demands even more.

The dynamical stability of the quasiequilibrium solutions obtained in
\cite{HMSU2008, UTGHSTY2016} is not yet known. If these objects are
dynamically unstable, do they undergo prompt collapse to black holes,
or do they evolve to significantly different, stable, axisymmetric
equilibria by rearranging their mass and/or angular momentum profiles?
If they are dynamically stable, their secular fate is still
unknown. Being nonaxisymmetric and rotating they will generate GWs,
which will radiate both energy and angular momentum. Will this lead to
delayed collapse to a black hole, or will it lead to the formation of
a Dedekind-like configuration, or something less exotic?  

In \cite{BSS00, SBS00} the \textit{dynamical} stability of
axisymmetric, differentially rotating stars (even including an initial
perturbation) has been studied numerically in GR and it was found that
they are stable against quasiradial collapse and bar-mode formation
for sufficiently small $\GB$.  GR enhances the dynamical bar-mode
formation since the critical value for $\beta=\GB_{\rm dyn}$ above
which the stars become dynamically unstable was found to be $\sim
0.24$, slightly less than the corresponding Newtonian value $0.27$ for
incompresible Maclaurin spheroids.  
A precise determination of the threshold for the dynamical instability,
the effects of stellar compactness on that, as well as the timescale of the 
persistence of the bar deformation have been studied in \cite{BPMR07,MBPR07}.
In \cite{SKE02, KE03, SKE03} linear stability analysis and simulations have
been performed to analyze the occurrence of the dynamical instability against 
nonaxisymmetric bar mode deformation for differentially rotating stars. It
was found that when differential rotation is high, the stars are dynamically
unstable even when $\GB$ was of order of $0.01$. This dynamical instability 
does not create spiral arms \cite{Centrella01, Saijo03, Ott05, Ou06, Paschalidis15, East16} 
or fragmentations, but drives the star into a 
quasi-stationary ellipsoid that emitts GWs.

The \textit{secular} bar-mode
instability induced by gravitational radiation with a polytropic
($\Gamma=2$) equation of state (EoS) in the 2.5 post-Newtonian
framework for rapidly rotating stars with $\GB\sim0.2-0.25$ has been
investigated in \cite{SK04}\footnote{The critical $\GB$ for instability in
Newtonian Maclaurin spheroids is $\GB_{\rm sec}\sim 0.14$, but
decreases in GR as the compaction increases \cite{SF98}}.  They 
tracked the evolution of the bar-mode
up until the final object was a deformed ellipsoid which was still
emitting GWs (therefore was not a Dedekind-like star). At the same
time the nonlinear developent of the secular bar-mode instability
using a stiffer EoS ($\Gamma=3$) and similarly including
post-Newtonian terms for the gravitational radiation reaction was
investigated in \cite{OTL04}. Although they were able to reach a
``Dedekind-like" state, this was destroyed after ten dynamical
times. According to the authors the reason could be either the
nonlinear coupling of various oscillatory modes in the star, or an
``elliptic flow'' instability which manifests itself when the fluid
flow is forced along elliptic streamlines.

In a previous work \cite{UTea2016} we computed for the first time
triaxial \textit{supramassive} neutron stars (uniformly rotating
models with rest-mass higher than the maximum rest-mass of a
nonrotating star, but lower than the maximum rest-mass when
allowing for maximal uniform rotation), by using a piecewise
polytropic EoS. In this work we perform the first evolutions of such
stars and try to investigate their stability and gravitational wave
content. Following \cite{UTea2016} we carefully construct five such
models: two \textit{normal} ones (uniformly rotating but not
supramassive) with compactions $0.1$ and $0.25$\footnote{These are the
  corresponding compactenesses of the spherical solutions.} adopting a
stiff, $\Gamma=4$, EoS, and three supramassive models with compactions
$0.23,\ 0.24,\ 0.26$ adopting a two-piece polytrope that has a soft
core. Although these EoSs are rather extreme, our goal is to prove a
matter of principle rather than focus on realistic EoSs.  For a single
polytrope a stiffer EoS can sustain a larger triaxial deformation, and 
hence the maximum mass of the triaxial star relative to that of the 
spherical star is expected to be larger. 
However, for the two-piece polytropic EoS, the maximum mass of the
triaxial star relative to the spherical counterpart
increases, even though the overall averaged stiffness of the EoS is 
softer.  If the mass difference between the maximum axisymmetric and
triaxial solutions is $\sim 10\%$ or less, then that implies that the
EoS of high density matter becomes substantially softer in the core of
neutron stars \cite{UTea2016}. 

We were able to follow the evolution of these objects for more than
twenty rotation periods, proving that they are \textit{dynamically
stable}. After an initial short period of time where junk radiation
in the initial data propagates away, the neutron star evolves along
quasiequilibrium states that satisfy the first law, $dM=\Omega
dJ$. Along this trajectory the orbital angular velocity remains
constant inside the neutron star, whose triaxial shape evolves toward
axisymmetry.  During this period the GW amplitude decreases
significantly, especially in the highly compact models. The question
that arises is: are we probing the secular fate of the stars or is
this clear monotonic amplitude decrease an artifact of numerical
dissipation.

We do not think that the decrease of the GW amplitude is due to
numerical viscosity. We performed a resolution study which did not
alter the main description above. We discuss the trigger for the
declining amplitude below.


\begin{table*}
\begin{tabular}{c|cccccc}
\hline
\hline
$\Gamma$   & $(P/\rho_0)_c$ & $\GR_c$ & $(\rho_0)_c$ & $M_0$ & $M$ & $M/R$  \\  
\hline
$4$        & $1.334$    & $0.004658$ & $0.003224$ & $2.882$ & $2.250$ & $0.3552$  \\  
$(4, 2.5)$ & $0.5674$   & $0.006175$ & $0.004536$ & $1.960$ & $1.657$ & $0.2871$  \\  
\hline
\hline
\end{tabular}
\caption{Characteristic quantities for the maximum mass spherical
  solutions of the two EoSs considered in this work. First line refers
  to simple polytrope models G4C010, G4C025, while second line to
  piecewise models pwC023, pwC024, pwC026. To convert to cgs units
  multiply mass, density and pressure by $1.989\times 10^{33}\ {\rm
    g}$, $6.173\times 10^{17}\ {\rm g/cm^3}$, and $5.548\times
  10^{38}\ {\rm g/(cm\ sec^2)}$, respectively.}
\label{tab:eos_param}
\end{table*}

If our models are imagined to sample bar-mode perturbations of an
axisymmetric configuration with $\GB>\GB_{\rm sec}$ then according to
well-known results \cite{SF98}, our stars should be secularly
unstable. We weren't able to find any growth of a bar-mode.  As in
\cite{DLSS04}, where evolutions of models with $\beta$ larger than
$\GB_{\rm sec}$ with an initial bar-mode perturbation were performed,
we find the decay of the initial perturbation.

Here we employ geometric units in which $G=c=M_\odot=1$, unless stated
otherwise. Greek indices denote spacetime dimensions (0,1,2,3), while
Latin indices denote spatial ones (1,2,3).

\section{Methods and physical parameters}
\label{sec:methods}

The numerical methods used here are those implemented in the \cocal{}
and \illinois{} codes, and have been described in great detail in our
previous works \cite{TU07, UT12, UTG12, TU13, TU15,
  TMGRU16, EFLSTB08, ELS10, EPLS12, PLES10, PLES11},
so we only summarize the most important features here.

\subsection{Initial data}
\label{sec:ID}

\begin{table}
\begin{tabular}{c|ccccc}
\hline
\hline
   & \multicolumn{5}{c}{Initial data models}  \\
   & G4C010 & G4C025 & pwC023 & pwC024 & pwC026  \\  
\hline
$\GR_0 (\times 10^{-3})$ & $1.005$    & $1.565$    & $1.902$    & $1.991$    & $2.226$   \\  
$\GR   (\times 10^{-3})$ & $1.019$    & $1.644$    & $2.065$    & $2.176$    & $2.477$   \\  
$R_x$                    & $7.677$    & $7.429$    & $7.774$    & $7.625$    & $7.266$   \\  
$R_z/R_x$                & $0.4727$   & $0.4957$   & $0.4977$   & $0.5015$   & $0.5108$   \\  
$R_y/R_x$                & $0.7500$   & $0.9063$   & $0.9219$   & $0.9375$   & $0.9688$  \\  
$e_z$                    & $0.8812$   & $0.8685$   & $0.8673$   & $0.8652$   & $0.8597$ \\  
$\Omega M$               & $0.01823$  & $0.08043$  & $0.07850$  & $0.08237$  & $0.09138$ \\  
$P$ (Period)             & $193.8$    & $138.3$    & $140.8$    & $137.7$    & $130.4$  \\  
$M$                      & $0.5623$   & $1.771$    & $1.760$    & $1.805$    & $1.896$   \\  
$M_0$                    & $0.5900$   & $2.012$    & $1.989$    & $2.047$    & $2.169$   \\  
$J/M^2$                  & $1.109$    & $0.8516$   & $0.8279$   & $0.8202$   & $0.8003$  \\  
$(M/R)_s$                & $0.1000$   & $0.2500$   & -          & -          & -          \\  
$M/R_x$                  & $0.07324$  & $0.2383$   & $0.2264$   & $0.2367$   & $0.2610$  \\  
$T/|W|$                  & $0.1543$   & $0.1773$   & $0.1666$   & $0.1661$   & $0.1633$   \\  
$I$                      & $10.81$    & $58.77$    & $57.46$    & $58.51$    & $59.70$ \\  
$\varepsilon_z$          & $0.2320$   & $0.05581$  & $0.02771$  & $0.0191$   & $0.006200$ \\  
$\GD M(\times 10^{-4})$  & $0.8237$   & $0.9893$   & $1.129$    & $1.063$    & $1.007$     \\  
VE$(\times 10^{-4})$     & $12.13$    & $5.463$    & $8.047$    & $7.753$    & $7.546$     \\  
\hline
\hline
   & \multicolumn{5}{c}{Quadrupole estimates}  \\
\hline
$\dot{E}(\times 10^{-8})$ & $2.846$  & $14.01$     & $2.918$    & $1.548$     & $0.2017$     \\  
$\dot{J}(\times 10^{-7})$ & $8.778$  & $30.84$     & $6.540$    & $3.391$     & $0.4186$     \\  
$rh/M(\times 10^{-3})$   & $14.63$   & $7.357$     & $3.441$    & $2.389$     & $0.7774$     \\  
\hline
\hline
   & \multicolumn{5}{c}{Timescales}  \\
\hline
$t_d/M$                  & $50$       & $10$       & $10$       & $10$       & $10$       \\  
$t_s/M$                 & $10^5$     & $10^5$      & $10^6$      & $10^6$      & $10^7$      \\  
\hline
\hline
\end{tabular}
\caption{Models G4C010, G4C025 have $\Gamma=4$ throughout, while
models pwC023, pwC024, pwC026 have $(\Gamma_1,\Gamma_2)=(4,2.5)$ and
are supramassive. Here $\GR_0$, $\GR$, $R_i$, $e_z=\sqrt{1-(R_y/R_x)^2}$,
$\Omega$, $P$, $M$, $M_0$, $J$, $(M/R)_s$, $T/|W|$, $I$,
$\varepsilon_z$, $\dot{E}$, $\dot{J}$, $h$, $t_d$, $t_s$ are the
rest-mass density, the total energy density, the coordinate radii, the
eccentricity with respect to the z-axis, the angular velocity, the
period, the ADM mass, the rest-mass, the ADM angular momentum, the
corresponding spherical compactness, the parameter $\GB$, the moment
of inertia, the ellipticity with respect to the z-axis
(Eq. \ref{eq:grell}), the luminosity, the angular momentum loss rate,
the GW maximum amplitude, the dynamical timescale, and the secular
timescale, respectively. 
To convert to geometric $G=c=1$ or cgs units, use the fact that 
$1=1.477\ {\rm km}=4.927\ {\rm \mu s} = 1.989\times 10^{33}\ {\rm g}$.}
\label{tab:id_param}
\end{table}

Our initial rotating star spacetimes posses a helical Killing vector,
$k^\GA$, \be k^\GA = t^\GA + \Omega \GP^\GA,
\label{eq:hkv}
\ee
whereby the fluid variables are Lie-dragged along $k^\GA$,
\be
\mathcal{L}_{\bf k}(hu_\GA) = \mathcal{L}_{\bf k}\GR_0 = \mathcal{L}_{\bf k} s = 0.
\label{eq:sym}
\ee
Here $u^\GA$ is the 4-velocity of the fluid, $\GR_0,\ h,\ s$ are the rest-mass density, enthalpy, and the 
entropy per unit rest-mass. We have $\GR_0 h=\GR+P$, where $\GR$ is the total energy density
and $P$ is the pressure. The 4-velocity
of the fluid will be along the helical Killing vector, i.e. there is a scalar $u^t$ such that 
\be
u^\GA = u^t k^\GA = u^t (t^\GA + v^\GA) = \GA u^t (n^\GA + U^\GA).
\label{eq:4vel}
\ee In the above, $v^i=\Omega\GP^i=\Omega (-y,x,0)$ is the velocity
with respect to the inertial frame, while $U^\GA$ is the spatial
velocity with respect to normal observers (those with 4-velocity
$n^\GA$). In the last equality, $\GA$ is the lapse function, that
normalizes the normal vector to the spacelike hypersurfaces which
foliate the spacetime, $n_\GA=-\GA\nabla_\GA t$.

For a perfect gas stress-energy tensor and an isentropic initial configuration the equations of motion yield a
first integral,
\be
\frac{h}{u^t} = \mathcal{E},
\label{eq:ei}
\ee where $\mathcal{E}$ is a constant. The two constants that appear
in our equations $\{\Omega,\mathcal{E}\}$ are determined via an
iterative scheme. For the
gravitational fields we use the Isenberg-Wilson-Mathews (IWM)
approximation \cite{I1980, WM1989} which assumes a flat conformal
metric and maximal slicing. The resulting five elliptic equations are
solved together with Eq. (\ref{eq:ei}) and coupled to a piecewise EoS
as described in \cite{HMSU2008,UTGHSTY2016}.

A number of diagnostics are used to describe the initial solutions and
explicit formulae are given in the appendix of \cite{UTGHSTY2016} and
will not be repeated here.  Since the IWM formulation is used, we have
that $\GG_{ij}=\GC^4 f_{ij}$, where $\GC$ is the conformal factor and
$f_{ij}$ the flat metric in spherical coordinates. The angular
momentum $J=J_{\rm ADM}$ [where $J_{\rm ADM}$ is the
  Arnowitt-Deser-Misner (ADM) angular momentum] is computed via a
surface integral at infinity or a volume integral over the spacelike
hypersurface. The kinetic energy is defined as $T:=\frac{1}{2}J\Omega$
(although we are not in axisymmetry we still use this formula because
it is gauge-invariant), and the gravitational potential energy is
defined as $W:=M_{\rm ADM}-M_{\rm P}-T$. Here $M_{\rm ADM}=M$ is the
(ADM) mass and $M_{\rm P}$ is the rest-mass plus internal energy of
the star (see e.g. \cite{BS10}).  These expressions are used then to
compute the rotation parameter $\beta$. Also the moment of inertial is
defined as $I:=J/\Omega$. As a measure of accuracy of the initial data
we provide two diagnostics: The first one is the difference between
the Komar and ADM mass,
\be \GD M = \frac{|M_{\rm K}-M_{\rm ADM}|}{M_{\rm K}} \,.
\label{eq:KmADM}
\ee
For stationary and asymptotically flat spacetimes $M_{\rm K}=M_{\rm
  ADM}$\footnote{Although for nonaxisymmetric systems the helical
  Killing vector (stationarity in the rotating frame) is incompatible
  with asymptotic flatness \cite{GS84}, one can define an approximate
  asymptotic region in which the gravitational wave energy is small
  compared with the total energy of the system. The same argument
  holds for the existence of the Komar mass that is associated with a
  timelike Killing field $t^\GA$.} \cite{Beig78}. The second
diagnostic is the relativistic virial equation (VE) \cite{GB94}.

The initial-data gravitational wave diagnostics involve the second
mass moments
\be I^{ij}:=\int_{\Sigma_t} \GR_0 u^\GA x^i
x^j dS_\GA
\label{eq:mij}
\ee
with $dS_\GA=\nabla_\GA t \sqrt{-g}d^3x$. In Appendix \ref{sec:quad}
we have derived some useful quantities such as the quadrupole
approximation for the luminosity and the gravitational wave amplitude,
that can be computed on a spacelike hypersurface in the presence of a
helical Killing vector. However, full GW output, including the
``junk'' radiation inherent in the initial data, is computed in full
GR as part of the integration of the field equations via the
Baumgarte-Shapiro-Shibata-Nakamura (BSSN) formalism~\cite{SN95,BS98}.

As in \cite{UTea2016} we employ the same ``benchmark" EoSs. The first
one is a simple $\Gamma=4$ polytrope, while the second is a
piecewise-polytropic EoS with two pieces and a soft core, where
$\{\Gamma_1,\Gamma_2\}=\{2.5,4\}$.  Characteristics of the maximum
mass solutions for spherical stars using these two EoSs are reported
in Table \ref{tab:eos_param}. The adiabatic constant $k$, is chosen so
that the value of the rest-mass becomes $M_0 = 1.5$ (in units of solar
mass) at the compactness $M/R = 0.2$. By choosing different values of
$k$ one can attain larger maximum masses.  A well-known fact that
relates the maximum masses of those models, is that a stiffer EoS can
sustain a larger maximum mass (see below). The same result holds for
the maximum masses of the axisymmetric solutions.  The values of
$\Gamma$ used are simply to prove a point of principle, rather than
address physical EoS parameters: stiffness is necessary in order for
these triaxial solutions to exist. A higher value of $\Gamma$
satisfies the necessary conditions for uniformly rotating triaxial
solutions to exist, and this is the main reason behind such a choice.
As discussed in \cite{UTea2016} the softening of the core enables us
to compute for the first time supramassive, triaxially deformed,
uniform rotating stars, without increasing further the maximum
polytropic exponent. This was made possible from the following
counter-intuitive fact which does not depend on the values of the
specific polytropic indices: Assume a simple (any $\Gamma>2.24$)
polytrope which in most cases does not support supramassive triaxial
solutions. Then consider a second two piece polytropic EoS
$\{\Gamma_1,\Gamma_2\}$, with $\Gamma_2=\Gamma$ and a soft core with
$\Gamma_1<\Gamma$. This second EoS is effectively softer than the
first. Thus one expects that the piecewise EoS does not exhibit 
triaxial solutions with mass larger than the maximum-mass spherical 
solution. This was proven not to be the case \cite{UTea2016}, and if the
relative difference between the maximum triaxial and axisymmetric masses 
is less than $10\%$\footnote{The maximum mass of triaxial solutions is 
always smaller than the maximum mass of axisymmetric ones.} 
it provides strong evidence of softening in the core of the compact object. 

In order to investigate the stability and gravitational wave signature
of such solutions we
consider five models, G4C010, G4C025, pwC023, pwC024, and pwC026 whose
characteristics are reported in Table \ref{tab:id_param}
\footnote{As we mentioned in the introduction all quantities reported
are in $G=c=M_\odot=1$ units. This means that if one wants to
convert mass to geometric $G=c=1$ units one has to multiply by
$1=1.477\ {\rm km}$. For the angular velocity $\Omega$, one divides
by $1=1.477\ {\rm km}$. Similarly to get $\Omega$ in cgs units again
one divides be $1=4.927\ {\rm \mu s}$.}.
The last three columns are
supramassive solutions while the others are normal ones. 
The triaxiality\footnote{Triaxiality is not used in any quantative way
in this paper. It can be defined in various ways, like $R_y/R_x,\ e_z$,
or $\varepsilon_z$ (see Table \ref{tab:id_param}) and signifies the departure 
from axisymmetry. In GW detection studies, triaxiality is measured by the 
ellipticity $\varepsilon_z$. Notice that the ellipticities of the models we consider here are 
larger than typical limits set by LIGO~\cite{Abbottetal17}. 
However, as isolated pulsars are dim and hard to find, there could exist
a population of undetected pulsars that LIGO has not probed, yet. } 
is larger for the first column model and diminishes as we move to more
compact stars. This means that the amplitude of the gravitational wave
will be larger for the first model and smaller for the last one.

\begin{table}
\begin{tabular}{lll}
\hline
\hline
$r_a=0$         &:& Radial coordinate where the radial grids start.       \\
$r_b=10^6$      &:& Radial coordinate where the radial grids end.     \\
$r_c=1.25$      &:& Radial coordinate between $r_a$ and $r_b$ where   \\
&\phantom{:}  & the radial grid spacing changes.   \\
$N_{r}=384$     &:& Number of intervals $\Dl r_i$ in $r \in[r_a,r_{b}]$. \\
$\Nrf=128$      &:& Number of intervals $\Dl r_i$ in $r \in[r_a,1]$. \\
$\Nrm=160$      &:& Number of intervals $\Dl r_i$ in $r \in[r_a,r_{c}]$. \\
$N_{\theta}=96$ &:& Number of intervals $\Dl \theta_j$ in $\theta\in[0,\pi]$. \\
$N_{\phi}=96$   &:& Number of intervals $\Dl \phi_k$ in $\phi\in[0,2\pi]$. \\
$L=12$          &:& Order of included multipoles. \\
\hline
\hline
\end{tabular}  
\caption{Summary of grid parameters used by \cocal{} to produce the five models.
Note that $\Nrf=128$ is the number of points across the largest star radius. }
\label{tab:grids_param}
\end{table}

\begin{table*}
\begin{tabular}{cccccccccc}
\hline
\hline
Model & $x_{\rm min}$ & $x_{\rm max}$ & $y_{\rm min}$ & $y_{\rm max}$ & $z_{\rm min}$ & $z_{\rm max}$ & 
Grid hierarchy & $dx$ & $N$  \\  
\hline
G4C010 & $-304$ & $304$ &  $-304$ &  $304$ &  $0$ &  $304$ & $\{9.5,19.0,38.0,76.0,152.0,304.0\}$ & $2.5\bar{3}$ & $96$  \\  
G4C025 & $-304$ & $304$ &  $-304$ &  $304$ &  $0$ &  $304$ & $\{9.5,19.0,38.0,76.0,152.0,304.0\}$ & $2.5\bar{3}$ & $93$  \\  
pwC023 & $-304$ & $304$ &  $-304$ &  $304$ &  $0$ &  $304$ & $\{9.5,19.0,38.0,76.0,152.0,304.0\}$ & $2.5\bar{3}$ & $98$  \\  
pwC024 & $-304$ & $304$ &  $-304$ &  $304$ &  $0$ &  $304$ & $\{9.5,19.0,38.0,76.0,152.0,304.0\}$ & $2.5\bar{3}$ & $96$  \\  
pwC026 & $-288$ & $288$ &  $-288$ &  $288$ &  $0$ &  $288$ & $\{9.0,18.0,36.0,72.0,144.0,288.0\}$ & $2.4$        & $96$  \\  
       & $-288$ & $288$ &  $-288$ &  $288$ &  $0$ &  $288$ & $\{9.0,18.0,36.0,72.0,144.0,288.0\}$ & $2.88$       & $80$  \\  
       & $-1152$& $1152$&  $-1152$&  $1152$&  $0$ &  $1152$& $\{9.0,18.0,36.0,72.0,144.0,288.0,576.0\}$ & $2.88$ & $96$  \\  
\hline
\hline
\end{tabular}
\caption{Grid parameters used for the evolution of each
model. Parameter $N$ corresponds to the number of points used to
cover the largest radius of the star. Parameter $dx$ is the step interval in the coarser
level. To convert to physical units multiply by $1=1.477\ {\rm km}.$}
\label{tab:evol_param}
\end{table*}

The models have been computed with the \cocal{} code, a second-order
finite-difference code whose methods are explained in
\cite{UT12,UTG12,TU13,TU15,TMGRU16}. For single compact objects it
emloys a single spherical patch $(r,\GU,\GP)$ with $r\in[r_a,r_b]$,
$\GU\in[0,\pi]$, and $\GP\in[0,2\pi]$, where $r_a=0$, $r_b=O(10^6 M)$,
and $M$ the total mass of the system (no compactification used).  The
grid structure in the angular dimensions is equidistant but not in the
radial direction.  The definitions of the grid parameters can be seen
in Table \ref{tab:grids_param}, along with the specific values used to
obtain the quasi-equilibrium solutions of this work.

\subsection{Evolution}
\label{sec:Evol}

For the evolution we use the \illinois{} code\footnote{We do not use
  {\tt IllinoisGRMHD}, which is the version of the code embedded in the
  Einstein Toolkit~\cite{EPHMS}.}, which solves the Einstein field
equations in the BSSN formalism \cite{SN95,BS98,BS10}.  The code is
built on the \cactus{} \cite{cactus} infrastructure and uses \carpet{} 
for mesh refinement, which allows us to focus numerical resolution on
the strong-gravity regions, while also placing outer boundaries at
large distances well into the wave zone for accurate GW extraction and
stable boundary conditions.  The evolved geometric variables are the
conformal metric $\tilde{\GG}_{ij}$, the conformal factor $\GP$,
($\GG_{ij}=e^{4\GP}\tilde{\GG}_{ij}$), the conformally-rescaled,
tracefree part of the extrinsic curvature, $\tilde{A}_{ij}$, the trace
of the extrinsic curvature, $K$, and three auxiliary variables
$\tilde{\Gamma}^i=-\pd_j\tilde{\GG}^{ij}$, a total of $17$ functions.
For the kinematical variables we adopt the puncture gauge conditions
~\cite{Baker06, Campanelli06,NRAR13}, 
which are part of the family of gauge conditions using an advective 
``1 + log'' slicing for the lapse, and a ``Gamma-driver'' for the
shift ~\cite{ABDKPST03}.
 
The equations of hydrodynamics are solved in conservation-law form adopting
high-resolution shock-capturing methods \cite{EPLS12, ELS10}.  The
primitive, hydrodynamic matter variables are the rest mass density,
$\GR_0$, the pressure $P$ and the coordinate three velocity
$v^i=u^i/u^0$. The enthalpy is written as $h=1+\GE+P/\GR_0$, and
therefore the stress energy tensor is $T_{\GA\GB}=\GR_0 hu_\GA u_\GB+
P g_{\GA\GB}$. Here, $\GE$ is the specific internal
energy\footnote{This should not be confused with the ellipticity
  $\varepsilon_z$.}.

To close the system an EoS needs to be provided and for that 
we follow \cite{PLES10,PLES11} where the pressure is decomposed as a
sum of a cold and a thermal part,
\be
P = P_{\rm cold} + P_{\rm th} = P_{\rm cold} + (\Gamma_{\rm th}-1)\GR_0 (\GE-\GE_{\rm cold})
\label{eq:pre}
\ee where \be \GE_{\rm cold} = -\int P_{\rm cold} d(1/\GR_0)
=\frac{k}{\Gamma-1}\GR_0^{\Gamma-1} + const.\ .  \ee Here $k,\Gamma$
are the polytropic constant and exponent of the cold part (same as the
initial data EoS) and $\Gamma_{\rm th}=5/3$. The constant that appears
in the formula above is zero for a single EoS, but takes different
values in a piecewise polytrope where one has to account for the
continuity of pressure at the join between the different pieces.

The grid structure used in these evolutions is summarized in Table
\ref{tab:evol_param}.  Typically we use six refinement levels with the
innermost level half-side length being approximately $\sim 1.25$ times
larger than the radius of the star in the initial data ($R_x$). We use
$240 \times 240 \times 120$ points for the innermost refinement level,
which means that we have approximately 190 points across the neutron
star largest diameter. (For the initial data construction we used 256 
points across the largest neutron star diameter.) 
For the innermost refinement level this implies a $\Delta x
\sim 0.0791\bar{6}=117\ {\rm m}$.  This number of points was necessary
in order to have accurate evolutions of such stiff EoS
($\Gamma=4$) which present a challenge for any evolution code.

For the last model pwC026 we have done two extra simulations, as the
compactness in this case was very high and the triaxiality very low. 
In this model the GW signal was very weak
($rh/M \sim 10^{-4}$) and therefore we wanted to corroborate our
findings by using different resolution and box size for the outer
boundary conditions.  On the last two lines of Table
\ref{tab:evol_param} the lower resolution simulation has the same
outer boundary distance ($288$) but $80$ points across the star
radius, while we have also a simulation with seven refinement levels
and the outer boundary at much larger distance ($1152.0$) than all
other cases.


\begin{figure*}
\begin{center}
\includegraphics[width=\columnwidth]{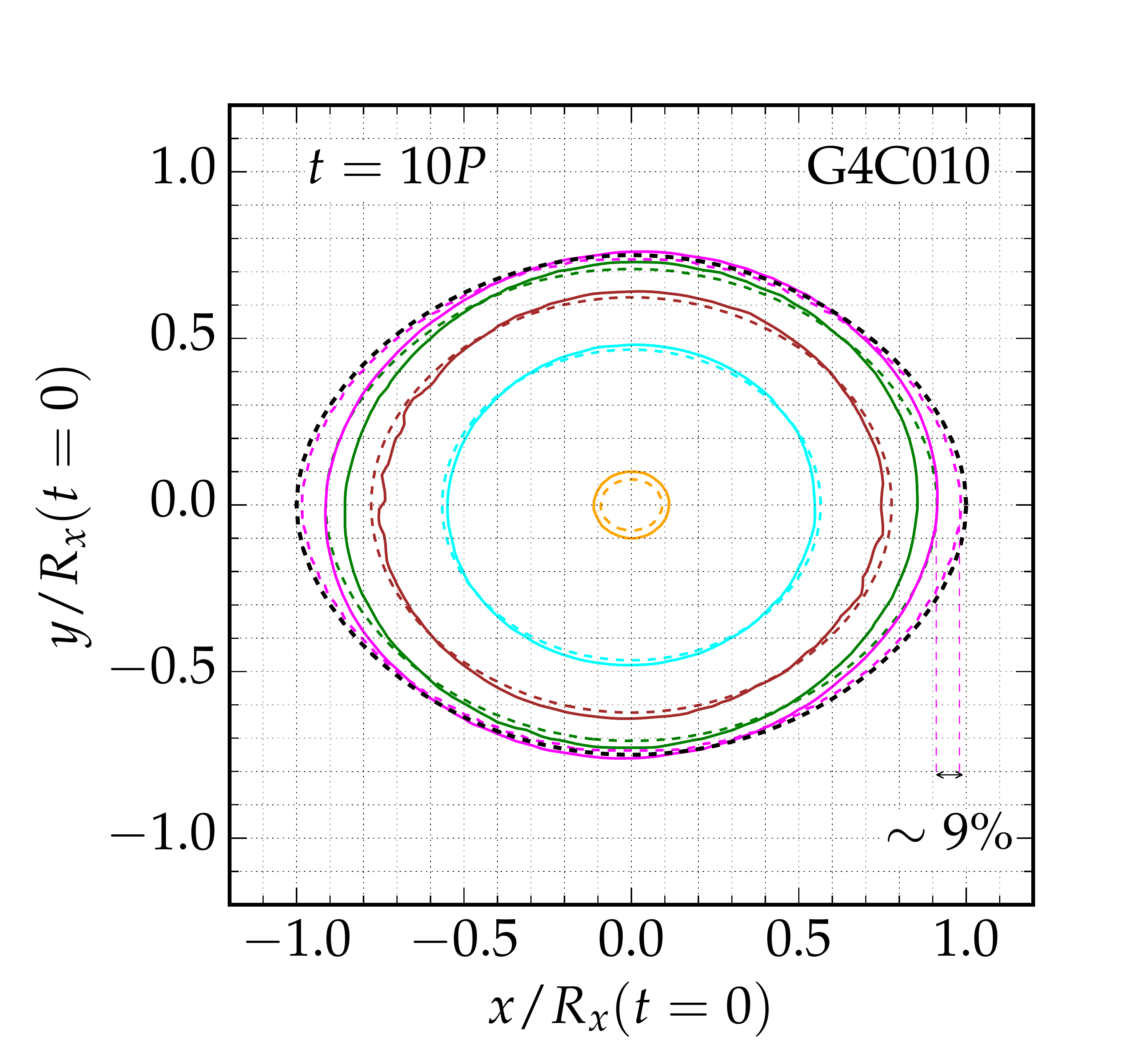}
\includegraphics[width=\columnwidth]{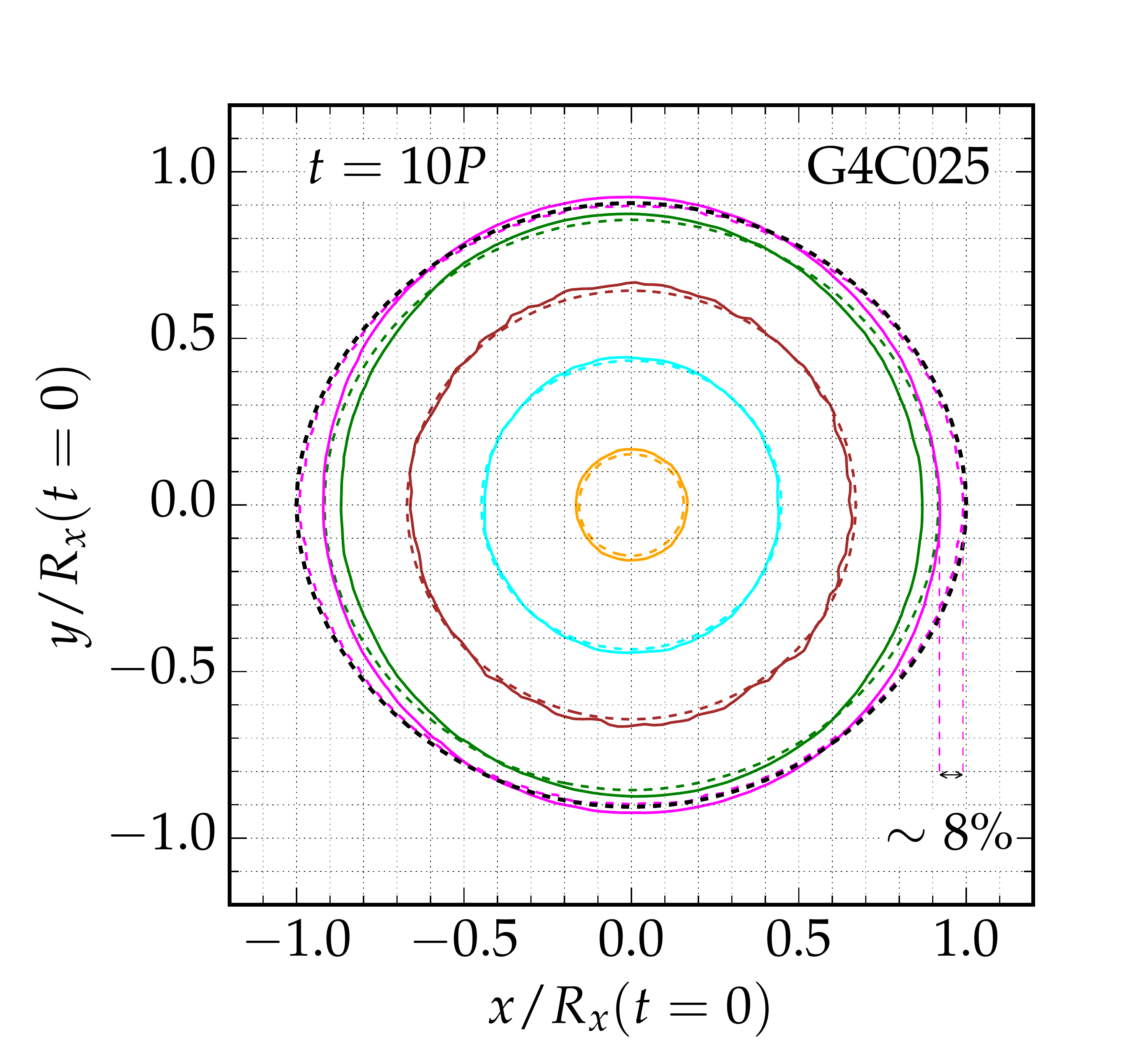}
\caption{Contour plots on the xy-plane of the rest-mass density for the normal models 
G4C010 and G4C025. Distances are normalized by the initial data radius along the x-axis $R_x(t=0)$.
The black dashed line signifies the initial data surface, while dashed color lines correspond to $t=0$
level lines of densities $\{0.2, 0.4, 0.6, 0.8, 1.0\}\times 10^{-3}$ 
for the G4C010 model. The same color 
but solid lines correspond to the same density levels after ten rotation periods. The contour plots of the 
G4C025 model correspond to $\{0.2, 0.5, 1.0, 1.3, 1.53\}\times 10^{-3}$.
To convert densities to cgs units multiply by $6.173\times 10^{17}\ {\rm g/cm^3}$.
}  
\label{fig:radii1}
\end{center}
\end{figure*}


\section{Results}
\label{sec:results}     

The main purpose of this work was to probe the stability of uniformly rotating, triaxial
configurations and estimate their gravitational wave emission. 
Although remnants from neutron star mergers have an asymmetric shape and can be dynamically
stable, they are differentially rotating. Is it possible for a triaxial star that
rotates \textit{uniformly} to be also dynamically stable? And if that is possible what is the
secular fate of this configuration?

\subsection{Dynamic stability}
\label{sec:dynsta}

\begin{figure*}
\begin{center}
\includegraphics[width=0.99\columnwidth]{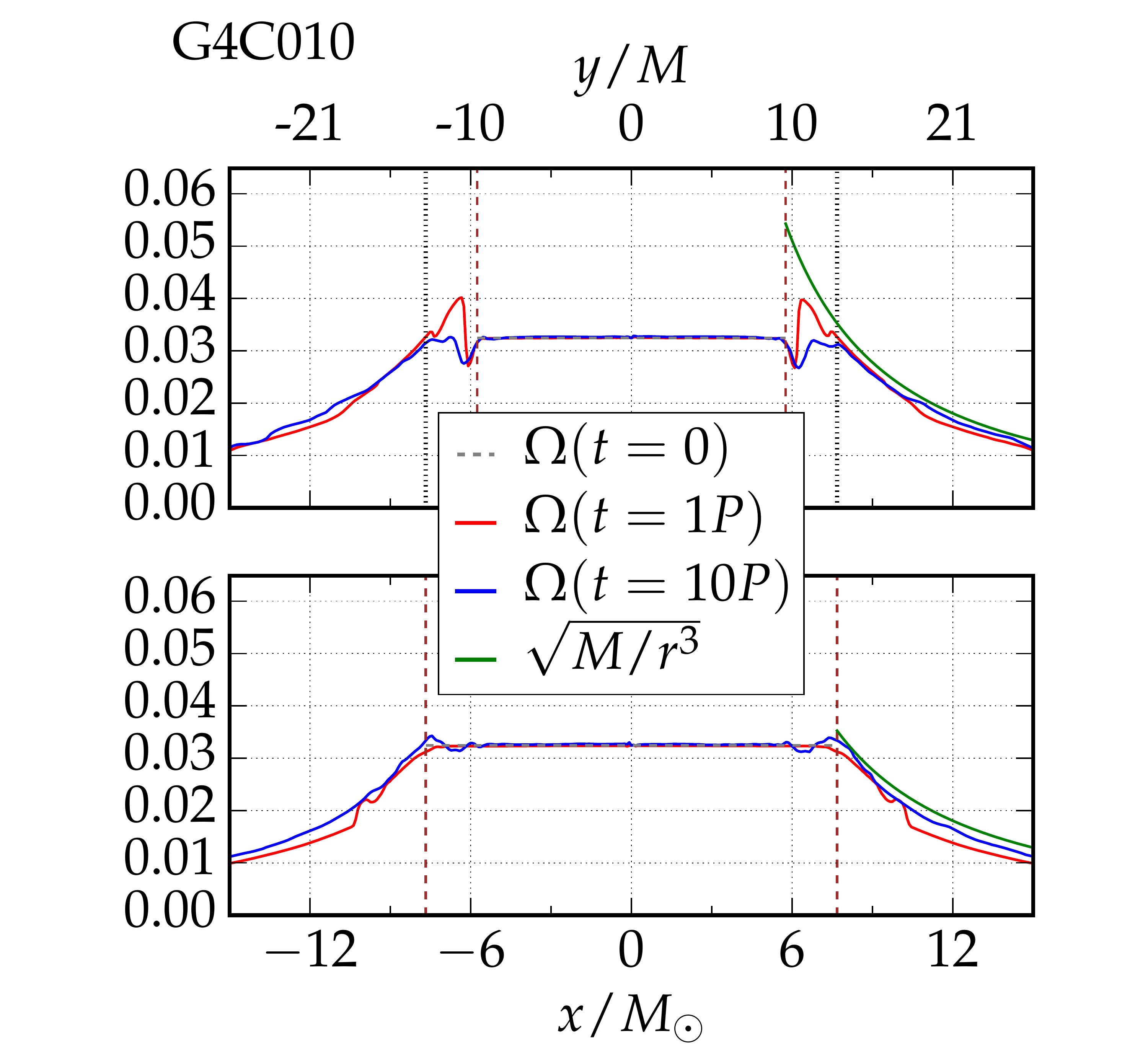}
\hskip 0.1cm
\includegraphics[width=0.99\columnwidth]{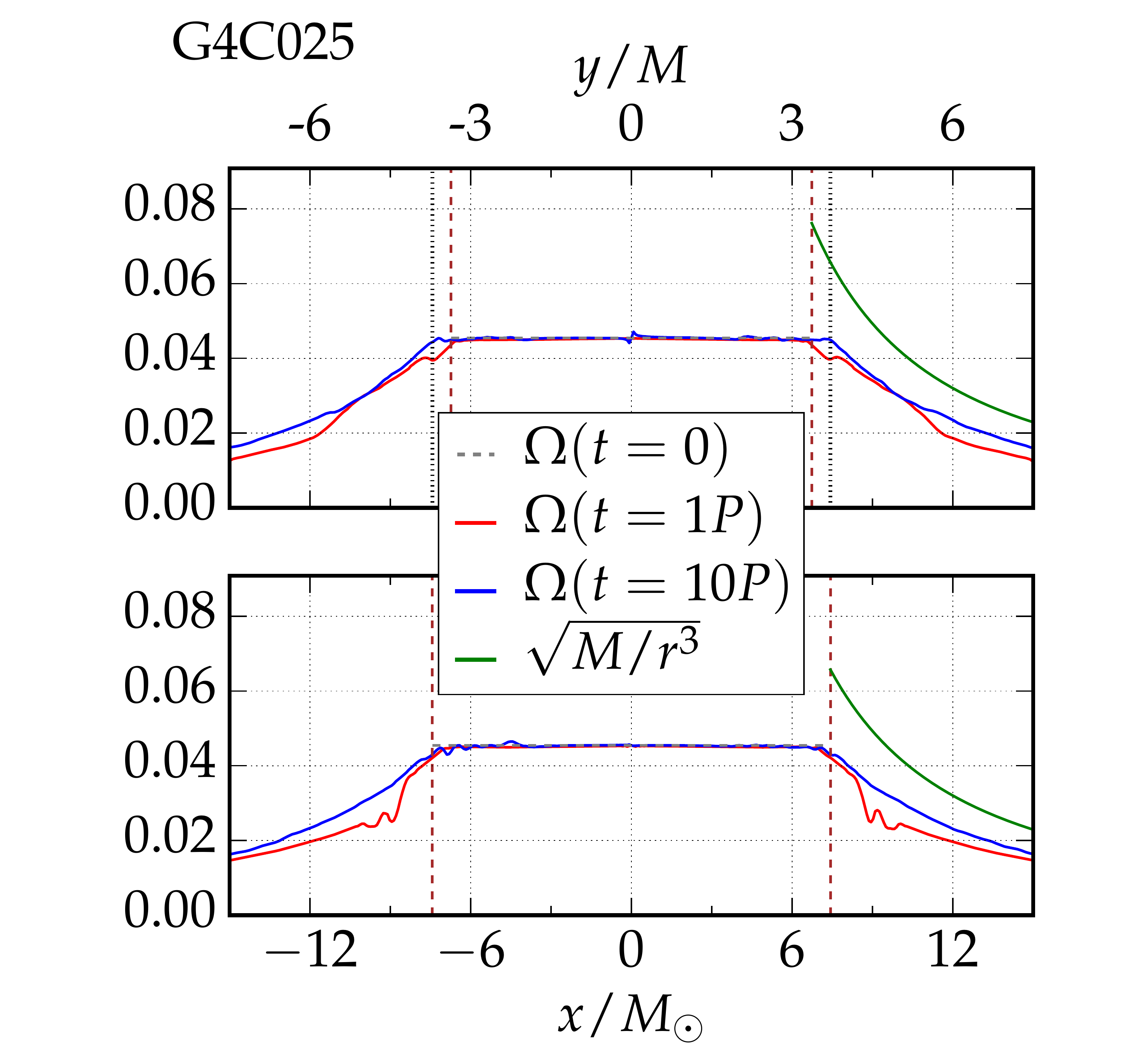}
\caption{Angular velocity profile across the x-axis (bottom) and the
  y-axis (top) for the normal models G4C010 and G4C025. The horizontal
  gray, dashed line corresponds to the initial data $\Omega$ and
  extends only in the interior of the star (this curve is difficult to
  see since it coincides with the red and blue curves inside the
  star).  Red and blue solid lines correspond to the angular velocity
  after one and ten rotation periods, while the green line is the
  Keplerian limit $(M/r^3)^{1/2}$.  Vertical brown dashed lines
  corresponds to the initial data radii along the x and y axes. 
  Vertical dotted gray lines on the top figures denote the initial radii
  along the x-axis. To convert $\Omega$ in cgs units divide by $4.927\ {\rm \mu s}$.
}  
\label{fig:ome}
\end{center}
\end{figure*}

The dynamical timescale for our stars is 
\begin{equation}
\frac{t_d}{M}\ \sim\ \frac{1}{\Omega M} \ \sim\ 
\left(\frac{M}{R}\right)^{-3/2}\, ,  
\label{eq:td}
\end{equation}
and values range from $\sim 10$ for the most compact cases to $\sim 50$ 
for  G4C010, the least compact model (see Table \ref{tab:id_param}). 
We find that all of the models considered are dynamically stable. Figure
\ref{fig:radii1} shows typical contour plots at $t=0$ (dashed colored
lines) as well as the same contour plots after ten rotation periods
(solid colored lines).
The black dashed line signifies the initial
data surface of the star in the xy-plane as calculated by \cocal{}.
Of particular importance are the lowest density contours at
$\GR_0=0.0002=9.16788\times 10^{-15}\ {\rm cm^{-2}}$ (magenta
colored).  The choice of this particular value can be considered as 
one of the largest densities that follow closely the initial data
profile (black dashed line).
By following the evolution
of this contour one can have an accurate picture of the surface of the
star.  After the junk radiation has propagated away the stars still
retain their triaxiality. But by $t=10\ P$, all contours tend to
circularize (the one of the highest density is initially circular).
All these contours contract in the x-direction and expand in the
y-direction. The amount of contraction/expansion diminishes as one
moves towards the center of the star. Thus the star becomes more
axisymmetric. After ten periods the x-axis has lost $9-8\%$ of its
length. For the supramassive models this picture still holds, although
since the ellipticities there are much smaller the amount of
contraction/expansion is somewhat diminished. For the most
supramassive model, pwC026, after ten rotation periods the decrease is
$\sim 4\%$ and the object is essentially axisymmetric.
While density contours are not gauge-invariant,
they yield a qualitative picture that agrees with the gravitational 
wave signature that we discuss in the next section.

The constant angular velocity profile is well preserved (Fig.
\ref{fig:ome}). The angular velocity across the x-axis (bottom panels)
and the y-axis (top panels) is plotted for the G4C010 and G4C025
models. Red curves correspond to $\Omega$ after one rotation period
while blue curves after ten rotation periods. Vertical brown dashed
curves denote the initial data star radii, and the green curve is the
Keplerian limit $\Omega_{\rm K}=(M/r^3)^{1/2}$.  The less compact the
star the closest to the Kepler limit is the ``atmospheric tail"
outside the surface of the star. Although the y-axis starts shorter
than the x-axis after ten rotation periods it has ``closed the gap"
and the two axes have essentially identical angular velocity profiles
(this gap is the space between vertical brown dashed and gray dotted
lines).  This effect is more evident in the G4C010 model but can be
clearly seen in the other most compact cases like G4C025.


\begin{figure*}
\begin{center}
\includegraphics[width=0.99\columnwidth]{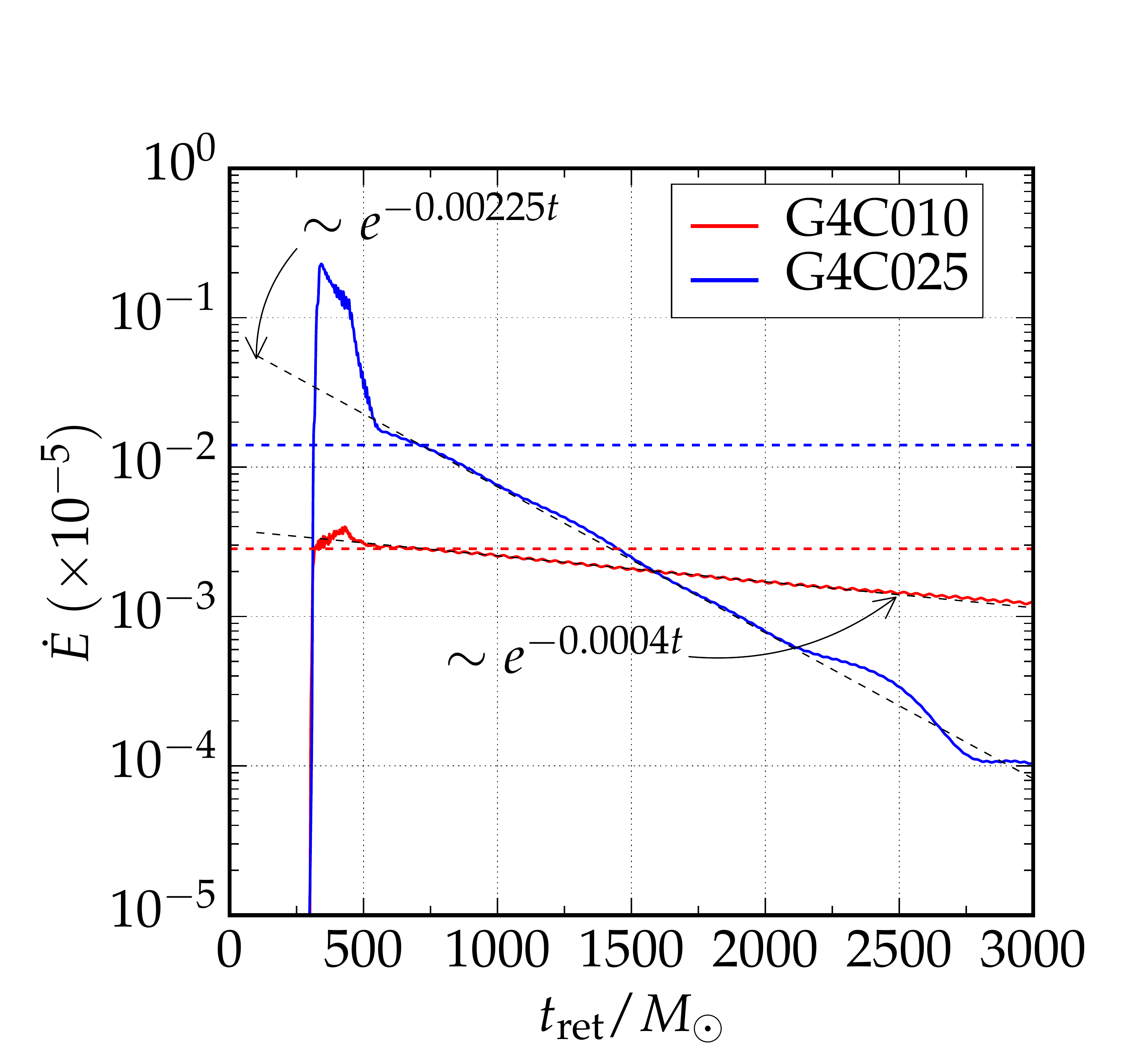}
\hskip 0.1cm
\includegraphics[width=0.99\columnwidth]{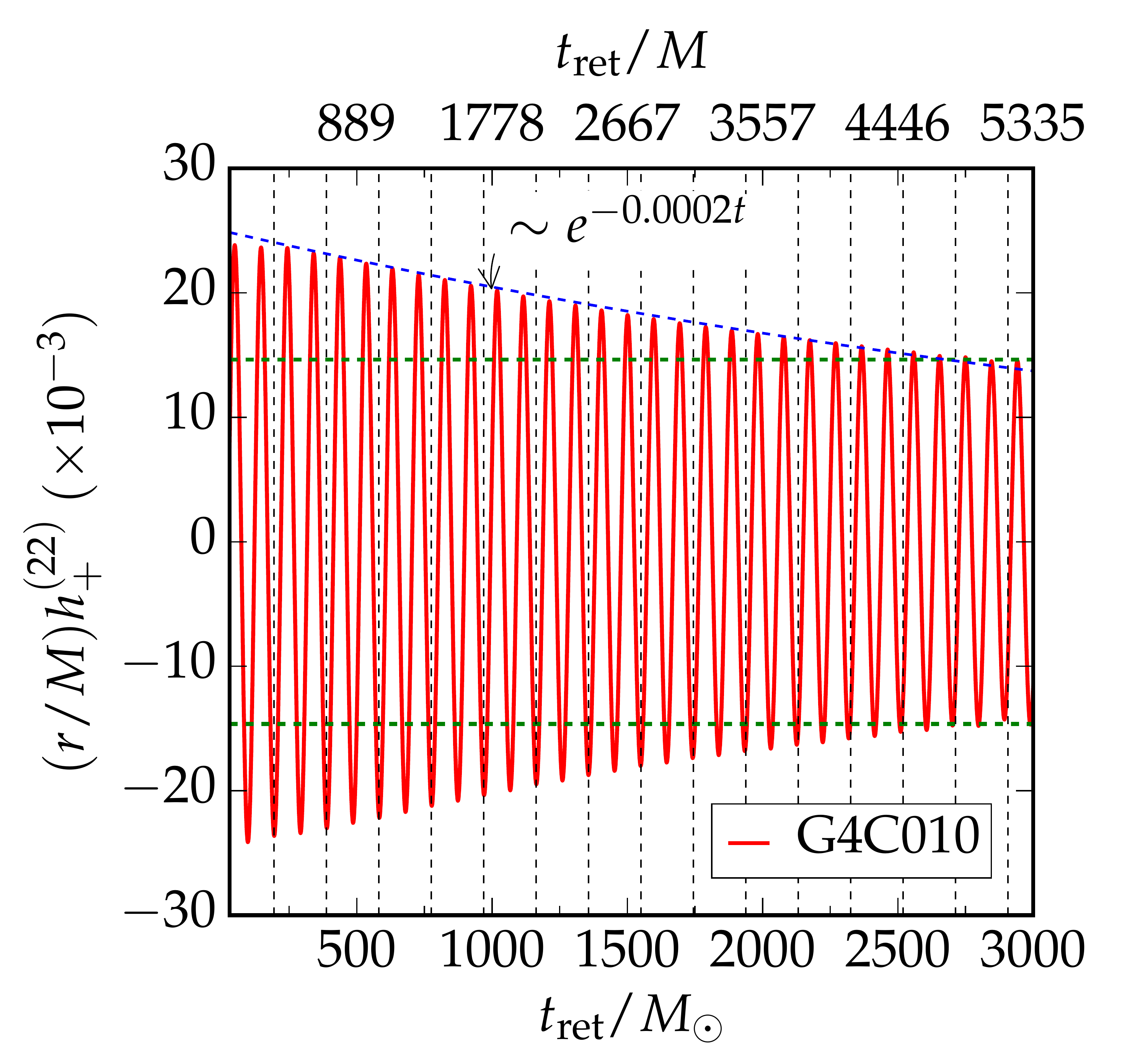}
\vskip 0.1 cm
\includegraphics[width=0.99\columnwidth]{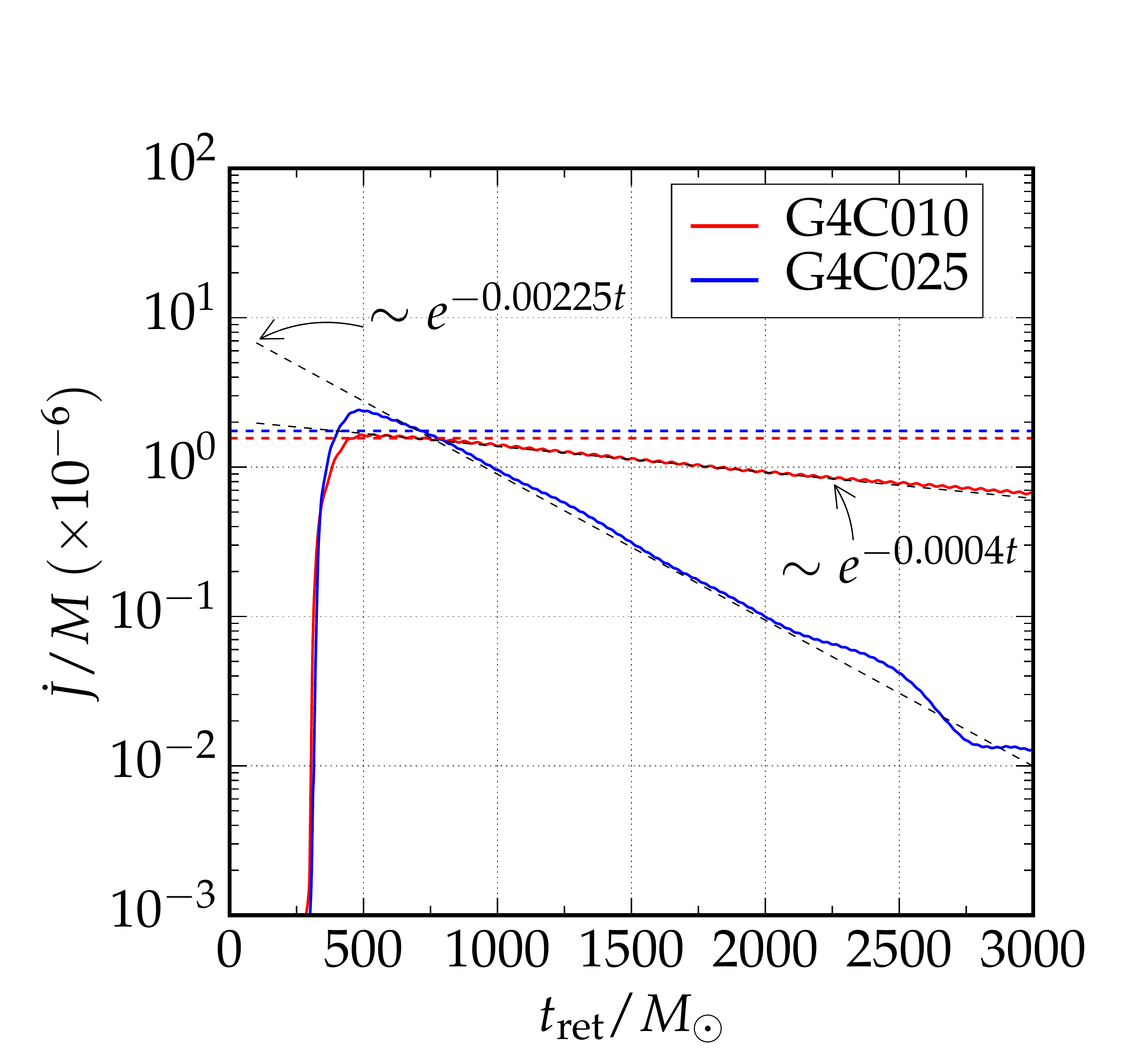}
\hskip 0.1cm
\includegraphics[width=0.99\columnwidth]{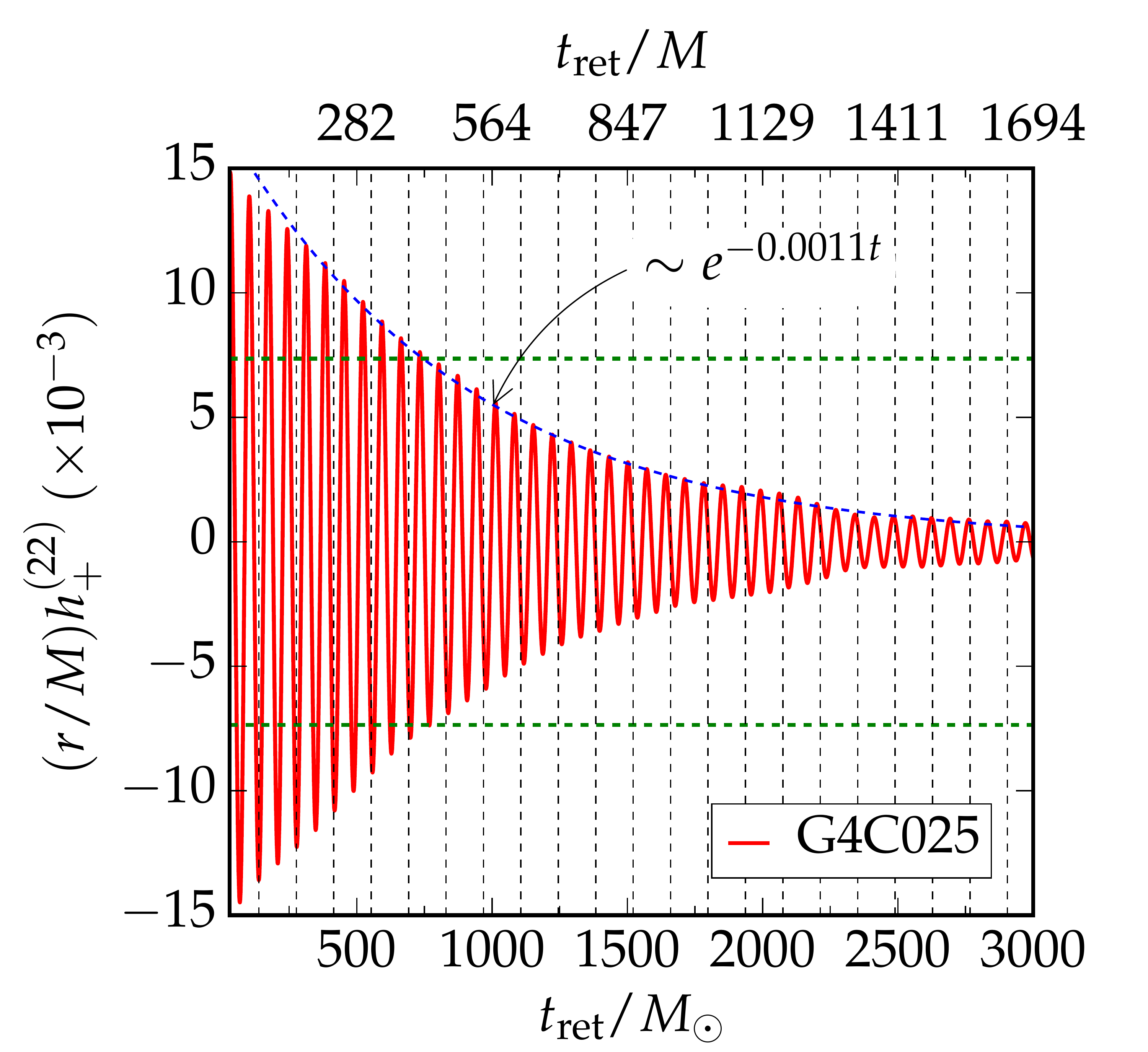}
\caption{All plots correspond to the normal models G4C010 and G4C025,
  and horizontal dashed lines are the initial data quadrupole
  estimates. Top left is gravitational wave power emitted; top right
  is the dominant $l=m=2$ mode of the gravitational wave strain for
  the least compact model G4C010, with vertical dashed lines
  corresponding to rotational periods; bottom left is the emitted
  angular momentum, and bottom right the strain for the G4C025
  model. Also denoted are exponential fitting curves. The GW
  timescales for the G4C010 and G4C025 models are $1/0.0002=8895\ M$
  and $1/0.0011=513\ M$ respectively. To convert $\dot{E}$
  and $\dot{J}/M$ to cgs units multiply by $1=3.629\times
  10^{59}\ {\rm ergs/sec}$ and $8.988\times 10^{20}\ {\rm ergs/g}$
  respectively.}
\label{fig:EJdot}
\end{center}
\end{figure*}

\begin{figure*}
\begin{center}
\includegraphics[scale=0.3]{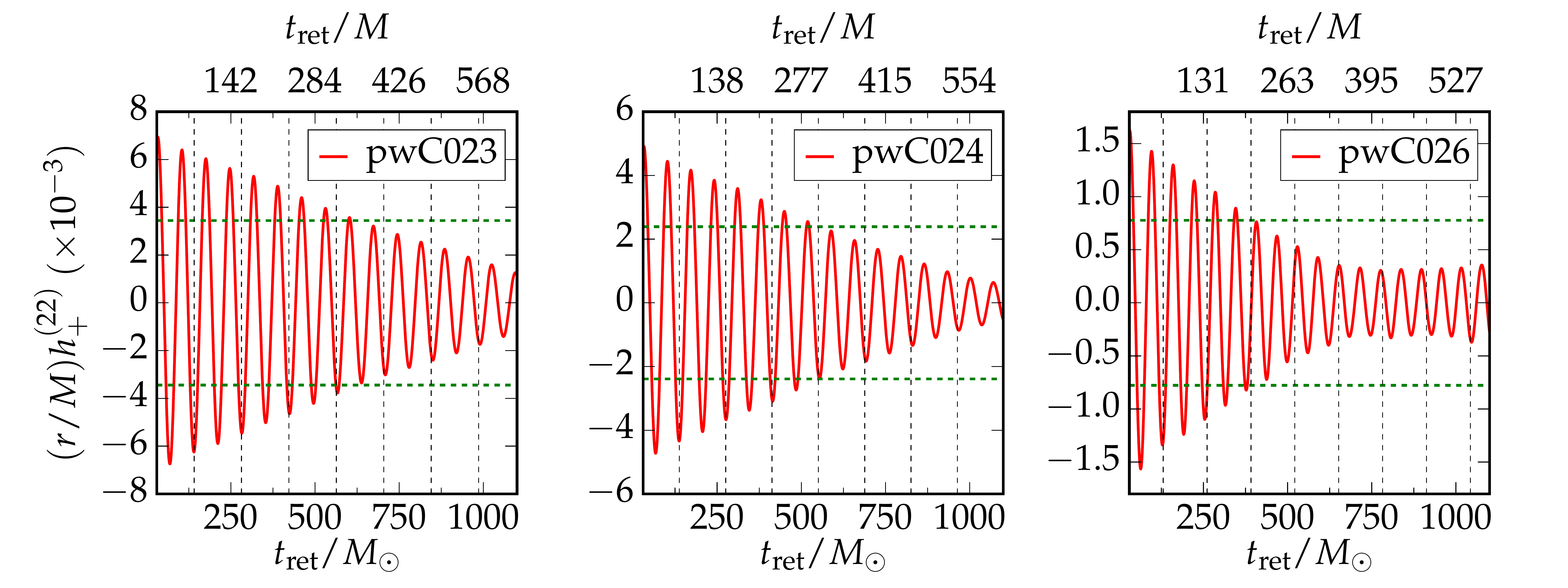}
\caption{GW strain for the supramassive models and the $l=m=2$ mode.
Vertical dashed lines correspond to rotational periods, while the horizontal dashed lines denote
the quadrupole approximation values.}  
\label{fig:GWpw}
\end{center}
\end{figure*}

\begin{figure*}
\begin{center}
\includegraphics[width=0.99\columnwidth]{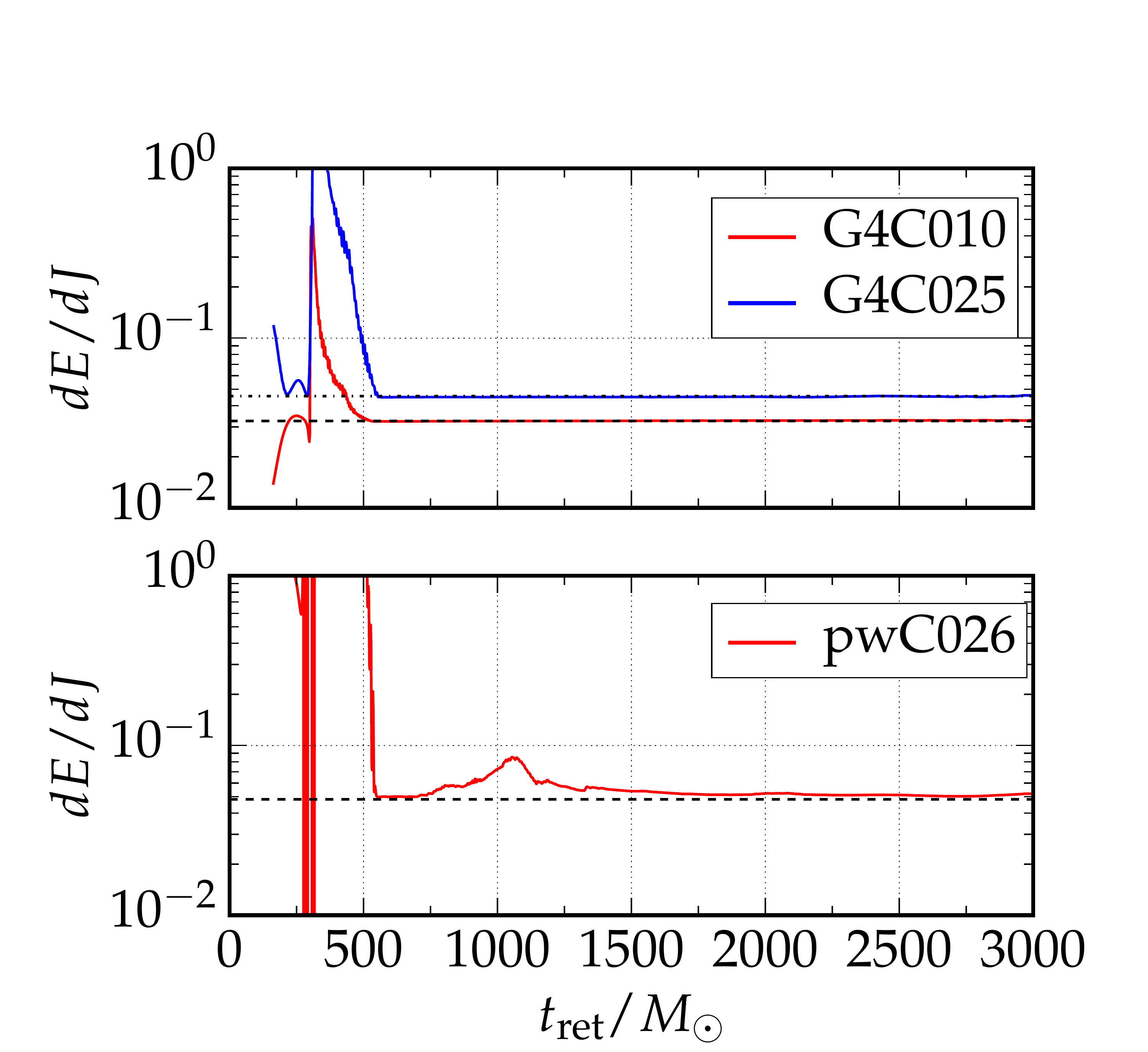}
\hskip 0.1cm
\includegraphics[width=0.99\columnwidth]{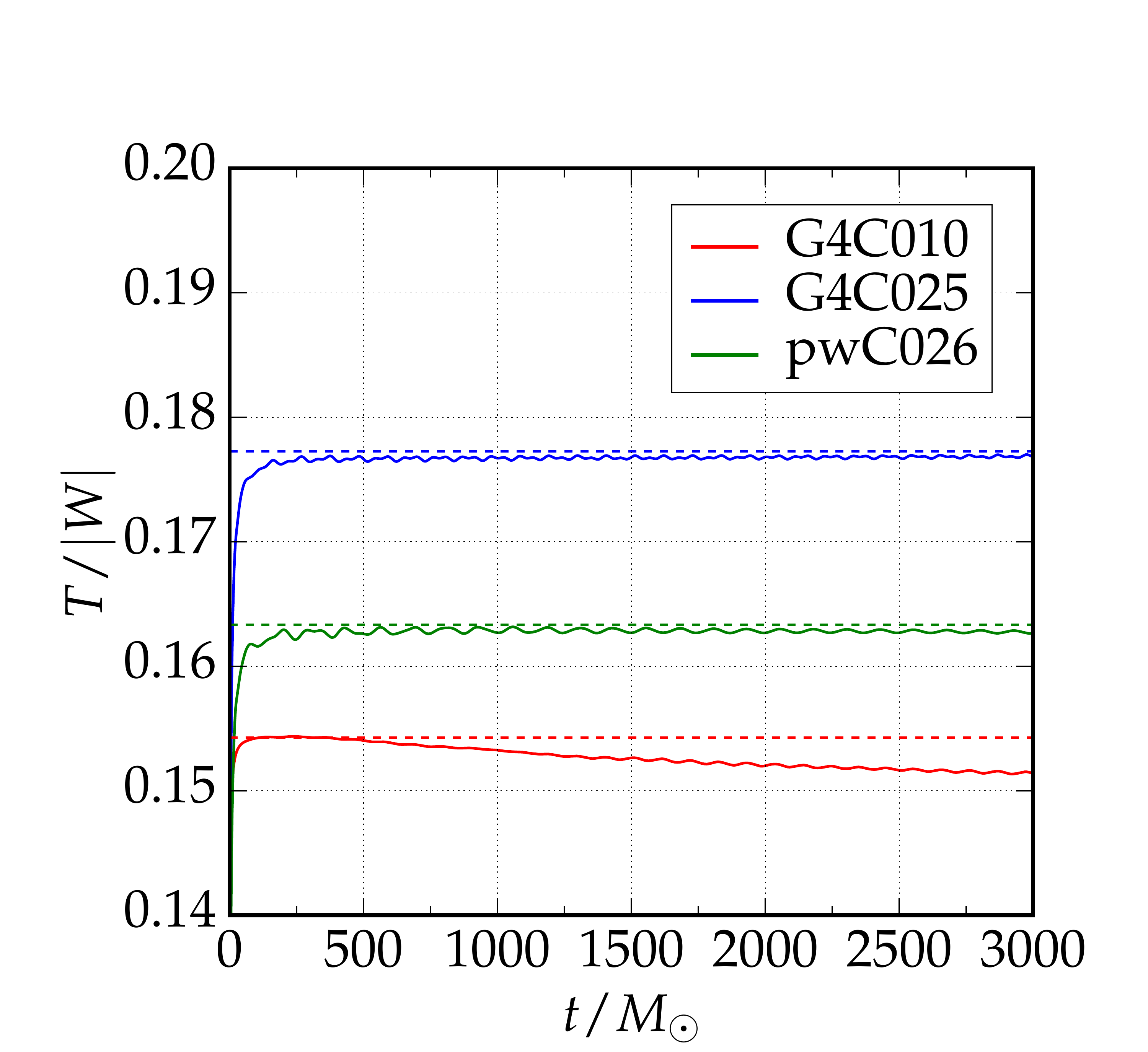}
\caption{Left plot: The first law for the triaxially deformed, uniformly rotating neutron stars (top are the
normal models G4C010 and G4C025, bottom is the most supramassive case pwC026). 
Dashed lines denote the corresponding initial data angular velocities. 
To convert $dE/dJ$ is cgs units divide by $4.927\ {\rm \mu s}$.
Right plot: Kinetic over gravitational potential energy again for the same models. Dashed lines 
denote the initial data values. }  
\label{fig:dEdJ}
\end{center}
\end{figure*}

\subsection{Secular fate}
\label{sec:secfate}

Although dynamic stability was straightforward to establish, that has
not been the case with secular stability. After evolving for more than
twenty rotational periods one can see in Fig.  \ref{fig:EJdot} the
major characteristics of GW emission. The frequency of
the dominant ($l=m=2$) GW mode is twice the rotational frequency, and
has amplitude approximately one-tenth of the
average value of a merging binary system. The
quadrupole-approximation prediction for the GW strain based on the
initial configurations is given by Eq. \ref{eq:h0}, and is shown in
the plots as a dashed horizontal green line. This approximate value
for the strain is about $50-60\%$ of the maximum amplitude found in
the evolution (see also Fig.~\ref{fig:GWpw}). The GW amplitude in the
more compact models (G4C025) experiences a more rapid decrease (almost
ten times more) than in the G4C010 case, which has the smallest
compactness ($0.1$). Similar behaviour is exhibited in the luminosity
and radiated angular momentum plots. In all cases, after an initial
period that lasts a little over $500\ M_\odot$ ($M_\odot=4.927\ {\rm
  \mu s}$) $\dot{E}$ and $\dot{J}$ intersect the predictions from the
quadrupole approximation based on the initial-data (in the plots these
are denoted by the horizontal blue and red dashed lines, Eqs
\ref{eq:qej}). However, $\dot{E}$ and $\dot{J}$ undergo exponential
decay in close agreement with the corresponding exponential decay in
the GW amplitude, i.e.,
\be 
\dot{E} \propto \dot{J} \propto (rh)^2 \, .  
\ee
In Fig. \ref{fig:EJdot} we denote the exponential fits for
all relevant functions.
The evolutionary path of these rotating objects occurs along
quasiequilibrium states as seen in the left panel of
Fig. \ref{fig:dEdJ}, which shows $dE=\Omega dJ$. After an initial
period of $\sim 500\ M_\odot$, this law is satisfied in all cases,
apart from a small perturbation at $1050\ M_\odot$ for the most
supramassive case pwC026. During that period the ratio of the kinetic
to gravitational potential energy remains essentially constant and
equal to the initial value (see right panel of Fig. \ref{fig:dEdJ}).

The thermal energy
generated by shocks was also measured in these simulations by inspecting the entropy parameter
$K:=P/P_{\rm cold}$, where $P_{\rm cold}=k\GR_0^\Gamma$.
With $\GE_{\rm th}=\GE-\GE_{\rm cold}=(K-1)(\Gamma-1)\GE_{\rm cold}/(\Gamma_{\rm th}-1)$, 
then $K>1$ implies shock heated gas 
\cite{ELSB09}. Since we don't have any mergers in our problem we didn't expect any
shocks, and this was the case for the bulk of the stars ($K\sim 1.0$). 

Although we clearly see that triaxially deformed stars evolve in a
quasiequilibrium manner towards axisymmetric objects, the key question
is whether this evolution is due mainly to GW emission or to a
hydrodynamical reconfiguration?  Using the exponential fitting
functions in Fig. \ref{fig:EJdot} we read off the GW decay
timescales. These are $5000M_\odot\sim 10^4\ M$ for the G4C010 and
$900M_\odot\sim 500\ M$ for the G4C025 models. The GW driven bar-mode
instability occurs for stars rotating with $\GB>\GB_{\rm sec}$ and
$\GB_{\rm sec}\approx 0.14$ in the Newtonian incompressible limit.
This value decreases in GR as the compactness increases \cite{SF98}.
The two models discussed here have $\GB=0.15, 0.18$ (see Table
\ref{tab:id_param}), and are thus greater the Newtonian critical value
$\GB_{\rm sec}$. The GW timescale is \cite{FS75} 
\be 
\frac{\GT_{\rm bar}^{\rm GW}}{M}\sim 2\times
      10^{-3}\left(\frac{M}{R}\right)^5(M\Omega)^{-6}(\GB-\GB_{\rm sec})^{-5} \, ,
\label{eq:GWts}
\ee 
where $\GB_{\rm sec}$ may be approximated by 
$\GB_{\rm sec}=0.115-0.048 M/M_{\rm max}^{\rm sph}$ \cite{MSB99}.  Here
$M_{\rm max}^{\rm sph}$ is the maximum spherical mass for the given
EoS. For our cases values are taken from Table \ref{tab:eos_param},
which imply that $\GT_{\rm bar}^{\rm GW} \sim 10^8\ M$ for the G4C010
and $\GT_{\rm bar}^{\rm GW} \sim 10^5 M$ for the G4C025 models,
respectively. 
Supramassive models pwC023, pwC024, pwC026 have timescales 
$\GT_{\rm bar}^{\rm GW} \sim 10^5 M$ too.
We also note that the GW timescales as calculated from
the crude quadrupole estimate, $t_s\sim T/|\dot{E}|$, and reported in
Table \ref{tab:id_param}, are in most
cases (except for G4C010) longer than the timescales obtained
from Eq. \ref{eq:GWts}.  Moreover, our configurations do not evolve
toward Dedekind-like ellipsoids as in the case of the bar-mode
unstable Newtonian configurations \cite{LS1995,Miller73}. It is
possible that the nonlinear growth of the instability is halted by
mode-mode coupling, as our triaxial configuration contains modes
beyond $m=2$.


Another possibility for a gravitational wave driven secular instability is the 
nonaxisymetric r-mode. For the $l=m=2$ mode the timescale is \cite{LOM98}
\be
\frac{\GT_{\rm r}^{\rm GW}}{M}\sim 10\left(\frac{M}{R}\right)^4(M\Omega)^{-6} \, ,
\label{eq:GWrmode}
\ee
which implies $\GT_{\rm r}^{GW} \sim 10^7\ M$ for the G4C010 and $\GT_{\rm bar}^{GW} \sim 10^5 M$ for the G4C025
respectively. These timescales  again are much longer than the timescales found numerically. 
Also, in this case the wave frequency $f_{\rm GW}=4/3 f_{\rm rot}$
therefore this possibility is also ruled out by our data, for which $f_{\rm GW}=2 f_{\rm rot}$.

Numerical viscosity, although nonzero, can in principle be
responsible. The presence of viscosity can damp a GW radiation
reaction-induced bar-mode instability \cite{LD77}, although it needs
to be properly tuned. However, we evolved with two different
resolutions and found no change in the behaviour which might have been
expected if numerical viscosity were significant.  Also we repeated
the calculation with the \whisky{} code \cite{Betal05, GR07, DAPRG13,
  RRG14} and got very similar results.  It may be that even a small
numerical viscosity over time is sufficient to damp the mode, given
the long timescale ($>> t_{\rm dyn}$) for GW emission. If we modeled
numerical viscosity by a turbulent viscosity $\nu\sim \GA R c_s\sim
\GA (R/M)^{1/2}$, where $c_s$ is the sound speed, then a damping time
of $10^4\ M$ associated with this would only require $\GA\sim 10^{-3}$
to be effective. Such a small value might go unnoticed by a modest
resolution study.  On the other hand, if viscosity were to dominate GW
dissipation, one still expects that the bar mode will be triggered
above $\GB=\GB_{\rm sec}$, since viscosity alone can drive the
instability, and the triaxiality would grow \cite{Shapiro04}, but
this is not observed.  Hence we conclude that although our triaxial
stars evolve towards axisymmetry, it is not the bar or r-mode secular
effects that are mainly responsible for this fate but rather a
hydrodynamical reconfiguration of the initial data.

\section{Discussion}

In this work we investigated the stability properties and
gravitational wave signatures of uniformly rotating, triaxial neutron
stars in GR.  Using the \cocal{} code we have constructed normal as
well as supramassive solutions in quasiequilibrium and we evolved them
for the first time with the \illinois{} code.

All five solutions that we considered are dynamically stable and
evolve secularly towards an axisymmetric configuration. Although we
monitored the evolution for more than twenty rotation periods we were
unable to probe the final (secular) fate of these stars, which is
orders of magnitude longer.  We coroborated our findings by using
different resolutions, placement of outer boundary conditions,
atmospheric treatments, and simulations with a different (\whisky{})
code.

According to \cite{SF98} a perturbed axisymmetric star with
$\GB>\GB_{\rm sec}$ will be secularly unstable and develop a bar
mode. In our case the initial models already contain a bar
perturbation and are rotating beyond the secular bar-mode instability
limit, but we found no further growth of a bar mode 
in the time frame of our simulations, which was shorter than the
predicted, theoretical secular timescale. On the contrary
we observed the decay of the star's triaxiality, which is in
accordance with previous investigations \cite{DLSS04}.

\acknowledgments 
We thank Cecilia Chirenti, Nikolaos Stergioulas and Enping Zhou, for 
useful discussions. This work was supported by 
NSF Grants PHY-1300903, PHY-1602536, PHY-1607449,
NASA Grants NNX13AH44G, NNX16AR67G (Fermi), and
JSPS Grants-in-Aid for Scientific Research (C) No. 26400274, and 15K05085.
VP gratefully acknowledges support from the Simons foundation. 
This work made use of the Extreme
Science and Engineering Discovery Environment (XSEDE), which is
supported by National Science Foundation grant number TG-MCA99S008.
This research is part of the Blue Waters sustained-petascale computing
project, which is supported by the National Science Foundation (awards
OCI-0725070 and ACI-1238993) and the State of Illinois. Blue Waters is
a joint effort of the University of Illinois at Urbana-Champaign and
its National Center for Supercomputing Applications. 
Some numerical computations were carried out on the XC30 system at the 
Center for Computational Astrophysics (CfCA) of the National Astronomical 
Observatory of Japan.

\appendix

\section{Quadrupole formulae in helical symmetry}
\label{sec:quad}

For an estimate of the GWs one can  compute the time 
derivatives of the quadrupole mass moments. Typically the quadrupole formula  reads
\be
h_{ij}(t,x^a) = \frac{2}{r} \left[ \frac{d^2 \qeq_{ij}^{TT}}{dt^2} \right]_{\rm ret} \,
\label{eq:hq}
\ee
where  $\qeq_{ij} := I_{ij}-\frac{1}{3}f_{ij}I_{kk}$, and  $\qeq_{ij}^{TT}$ 
is the transverse traceless reduced quadrupole moment \cite{MTW}. 
The second time derivatives 
are computed at a retarded time. The corresponding gravitational wave luminosity and the angular momentum
carried away per unit time are 
\be
\frac{dE}{dt}=\frac{1}{5} \langle \dddot{\qeq}_{ij}\dddot{\qeq}_{ij} \rangle , \quad
\frac{dJ^i}{dt}=\frac{2}{5} \GE^{ijk} \langle \ddot{\qeq}_{ja}\dddot{\qeq}_{ka} \rangle \, , 
\label{eq:qej}
\ee
where $\langle\cdot\rangle$ denote an average over several wavelengths.
In full dynamical spacetimes there is no unique definition of the quadrupole moment but 
typically one uses Eq. (\ref{eq:mij}) as a generalized integral over the hypersurface $\Sigma$, \cite{SS03}
which can be thought as an Euclidean integral over a weighted density 
$\GR_\ast=\GR_0 u^t \sqrt{-g}$. Its time derivative
\be
\frac{d}{dt} I_{ij} = \int_\Sigma \GR_0 u^\GA (x_i v_j + x_j v_i) dS_\GA \, ,
\ee
can be obtained by using the conservation of rest mass $\pd_t \GR_\ast+\pd_i(\GR_\ast v^i)=0$, 
and integration by parts \cite{FE90}. 

Another way to obtain the same result is to employ the transport theorem that says that for
any density $\GR_\ast$ that satisfies the above continuity equation and any function
$Q(t,x^i)$, we have
\be
\frac{d}{dt} \int_{V_t} \GR_\ast Q dV = \int_{V_t} \GR_\ast \frac{DQ}{Dt} dV \, ,
\label{eq:tt}
\ee where $\frac{DQ}{Dt}=\pd_tQ + v^i\pd_i Q$ is the Lagrangian
derivative of $Q$.  For a fluid velocity $v^i=\Omega \phi^i$, we have
$DQ/Dt =\mathcal{L}_{\bf k}Q$, and thus we can write a fully 4-dim
version of the classical theorem as
\be \frac{d}{dt} \int_{\Sigma_t} Q
\GR_0 u^\GA dS_\GA = \int_{\Sigma_t} \mathcal{L}_{\bf k}Q \GR_0 u^\GA
dS_\GA \, .
\label{eq:tt4d}
\ee
A straightforward proof of Eq. \ref{eq:tt4d} can be obtained if we consider 
$f(t)=\int_{\Sigma_t} Q \GR_0 u^\GA dS_\GA$. Let $\Sigma=\Sigma_0$ and $\Sigma_t=\GC_t(\Sigma)$, 
where $t^\GA$ is the generator of the diffeomorphism family $\GC_t$. Then
\begin{eqnarray*}
f'(0) & = & \lim_{t\rightarrow 0} \frac{1}{t} 
\left\{ \int_{\Sigma_t} Q \GR_0 u^\GA dS_\GA  -  \int_{\Sigma} Q \GR_0 u^\GA dS_\GA\right\}  \\
& = & \lim_{t\rightarrow 0} \frac{1}{t} 
\left\{ \int_{\Sigma} \GC_{-t}(Q\GR_0 u^\GA dS_\GA)  -  \int_{\Sigma} Q \GR_0 u^\GA dS_\GA\right\}  \\
& = & \int_{\Sigma} \lim_{t\rightarrow 0}\frac{1}{t} 
\left\{\GC_{-t}(Q \GR_0 u^\GA dS_\GA)  - (Q \GR_0 u^\GA dS_\GA) \right\} \\
& = & \int_{\Sigma} \mathcal{L}_{\bf t}(Q \GR_0 u^\GA dS_\GA) =
      \int_{\Sigma} \mathcal{L}_{\bf t}(Q \GR_0 u^t \sqrt{-g}) d^3x   \\
& = & \int_{\Sigma} \mathcal{L}_{\bf{k}}(Q \GR_0 u^t \sqrt{-g}) d^3x - 
\Omega \int_{\Sigma} \mathcal{L}_{\Bphi}(Q \GR_0 u^t \sqrt{-g}) d^3x \\
& = & \int_{\Sigma} \mathcal{L}_{\bf{k}}(Q \GR_0 u^t \sqrt{-g}) d^3x - 
\Omega \int_{\Sigma} D_i(Q \GR_0 u^t \GA \GP^i) dS \\
& = & \int_{\Sigma} \mathcal{L}_{\bf{k}}(Q) \GR_0 u^\GA dS_\GA \, .
\end{eqnarray*}
To obtain the last line we converted the second integral in the
previous line over a divergence, to a surface integral that vanishes,
and also used the continuity equation in the form $\mathcal{L}_{\bf
  k}(\GR_0 u^t \sqrt{-g})=0$.

For the computation of Eq. (\ref{eq:qej}) we need to compute the third material derivatives of $x^i x^j$.
We denote by $\GP^i=(\GP^A,0)$ where capital letters take values in $\{1,2\}$. Then $\GP^A=-\GE^{AB}x_B$
and the nonzero components are
\begin{eqnarray*}
\frac{Dx^A}{Dt} & = & \Omega \GP^A := v^A   \\
\frac{Dv^A}{Dt} & = & -\Omega^2 x^A := a^A  \\
\frac{Da^A}{Dt} & = & -\Omega^3 \GP^A   \, .    \\
\end{eqnarray*}
Setting $\varpi^i=(x^A,0)$ we have
\begin{eqnarray*}
\dot{I}^{ij}(0)   & = & \Omega\int_{\Sigma} \GR_0 u^\GA (x^i \GP^j+ x^j \GP^i) dS_\GA \,, \\
\ddot{I}^{ij}(0)  & = & -\Omega^2\int_{\Sigma} \GR_0 u^\GA (\varpi^i x^j - 2\GP^i \GP^j + x^i \varpi^j) dS_\GA \,,  \\
\dddot{I}^{ij}(0) & = & -\Omega^3\int_{\Sigma} \GR_0 u^\GA (\GP^i x^j+ 3\varpi^i \GP^j + 3\GP^i \varpi^j + x^i \GP^j) dS_\GA . 
\end{eqnarray*}

Using the derivatives of the multiple moments above one can compute the luminosity
or the angular momentum radiated from Eq. \ref{eq:qej}. For
the GW strain, assuming rotation around the z-axis, we have
\begin{eqnarray*}
[h_{AB}] & = & \frac{2}{r}
\begin{bmatrix}
(\ddot{I}_{11} -\ddot{I}_{22})/2 &  \ddot{I}_{12}                     \\
\ddot{I}_{21}                    & -(\ddot{I}_{11} -\ddot{I}_{22})/2  \\
\end{bmatrix}   \\
         & = & 
\begin{bmatrix}
h_{+}      & \ h_{\times}  \\
h_{\times} & -h_{+}
\end{bmatrix}      \, .
\end{eqnarray*}

For the case of an exact triaxial ellipsoid
the two elliptical polarization modes for head on observation along the z-axis, we set
\be
h_{(+,\times)}=\frac{4\Omega^2}{r}(I_1-I_2)\ (\cos(2\Omega t), \sin(2\Omega t)) \, ,
\label{eq:h0}
\ee
where $I_k$ are the principal moments of inertia.
Then the emitted power and angular momentum will be,
\begin{eqnarray}
|\dot{E}| \ =\ \frac{32}{5} (I_1-I_2)^2  \Omega^6 , \\
|\dot{J}| \ =\ \frac{32}{5} (I_1-I_2)^2  \Omega^5 .
\label{eq:lum}
\end{eqnarray}

A parameter which is often mentioned is called ellipticity of the source is defined as
$\varepsilon :=|I_1-I_2|/I_3$. Although there is no rigorous
counterpart in GR we can generalize  as 
\be
\varepsilon_z := \frac{|I_{11}-I_{22}|}{I_{11}+I_{22}} \, .
\label{eq:grell}
\ee
This is the quantity that is reported in Table \ref{tab:id_param}.

\bibliographystyle{apsrev4-1}

\begin{thebibliography}{200}

\bibitem{abbott2016a} 
B. P. Abbott, R. Abbott, T. D. Abbott, \textit{et al.}, \PRL{116}{061102}{2016}. 



\bibitem{EatH}
https://einsteinathome.org/

\bibitem{Aasietal2013}
J.Aasi \textit{et al.}, \PRD{87}{042001}{2013}.

\bibitem{Papaetal2016}
M. A. Papa \textit{et al.}, \arxiv{1608.08928}.

\bibitem{CA}
J.D.E. Creighton and W.G. Anderson, Gravitational-Wave Physics and Astronomy: An Introduction
to Theory, Experiment, and Data Analysis, Wiley-VCH, 2011.

\bibitem{A2010}
N. Andersson, V. Ferrari, \textit{et al.}, \GRG{43}{409}{2011}. 

\bibitem{K03}
K. Kokkotas, Oscilattions and instabilities of relativistic stars, in Gravity Astrophysics and 
Strings 2002, editors P.P.Fiziev and M.D. Todorov, St. Kliment Ohridski University Press, Sofia, 2003.

\bibitem{ST83}
S. Shapiro and S. Teukolsky, Black Holes, White Dwarfs, and Neutron Stars. Wiley, New York (1983).

\bibitem{HJA2006}
B. Haskell, D. I. Jones, and N. Andersson, \MNRAS{373}{1423}{2006}.

\bibitem{FS14}
J. L. Friedman and N. Stergioulas, Instabilities of relativistic stars
General Relativity, Cosmology and Astrophysics, Fundamental Theories of Physics, Volume 177. 
Springer International Publishing Switzerland, 2014.

\bibitem{BR2016}
L. Baiotti and L. Rezzolla, \arxiv{1607.03540}

\bibitem{LS1995}
D. Lai and S. L. Shapiro, \AJ{442}{259}{1995}.

\bibitem{PO2011}
A. L. Piro and C. D. Ott, \AJ{736}{108}{2011}.

\bibitem{FS2013}
J. L. Friedman and N. Stergioulas, ``Rotating Relativistic Stars". 
Cambridge Monographs on Mathematical Physics. Cambridge University Press (2013)

\bibitem{PS2016}
V. Paschalidis and N. Stergioulas, \arxiv{1612.03050}

\bibitem{N1997}
T. Nozawa, Ph.D. thesis, University of Tokyo, 1997.

\bibitem{HMSU2008}
X. Huang, C. Markakis, N. Sugiyama, K. Ury\=u, \PRD{78}{124}{2008}.

\bibitem{UTGHSTY2016} 
K. Ury\=u, A. Tsokaros, F. Galeazzi, H. Hotta, M. Sugimura, K. Taniguchi and S. I. Yoshida, \PRD{93}{044056}{2016}.

\bibitem{James64}
R. A. James, \AJ{140}{552}{1964}.

\bibitem{BFG1996}
S. Bonazzola, J. Frieben, and E. Gourgoulhon,  \AJ{460}{379}{1996}.

\bibitem{Ch69}
S. Chandrasekhar, ``Ellipsoidal Figures of Equilibrium'',  New Haven: Yale University Press.

\bibitem{SZ1996}
S.L. Shapiro and S. Zane, \AJ{460}{379}{1996}.

\bibitem{BFG1998}
S. Bonazzola, J. Frieben, and E. Gourgoulhon,  \ASAS{331}{280}{1998}.

\bibitem{GV2002}
T. Di Girolamo and M. Vietri, \AJ{581}{519}{2002}.

\bibitem{Saijo2006}
M. Saijo, and E. Gourgoulhon, \PRD{74}{084006}{2006}.

\bibitem{RG2002}
D. Gondek-Rosinska, and E. Gourgoulhon,  \PRD{66}{044021}{2002}; 


\bibitem{SL1996}
D. Skinner, and L. Lindblom, \AJ{461}{920}{1996}.

\bibitem{YE1997} 
S. Yoshida, and Y. Eriguchi, \AJ{490}{779}{1997}.  


\bibitem{BSS00}
T. W. Baumgarte, S. L. Shapiro, and M. Shibata, \AJ{528}{L29}{2000}

\bibitem{SBS00}
M. Shibata, T. W. Baumgarte, and S. L. Shapiro, \AJ{542}{453}{2000}.

\bibitem{BPMR07}
L. Baiotti, R. De Pietri, G. Mario Manca, and L. Rezzolla, \PRD{75}{044023}{2007}.

\bibitem{MBPR07}
G. Mario Manca, L. Baiotti, R. De Pietri and L. Rezzolla, \CQG{24}{S171}{2007}.

\bibitem{SKE02}
M. Shibata, S. Karino and Yoshiharu Eriguchi, \MNRAS{334}{L27}{2002}.

\bibitem{KE03}
S. Karino and Y. Eriguchi, \AJ{592}{1119}{2003}.

\bibitem{SKE03}
M. Shibata, S. Karino and Y. Eriguchi, \MNRAS{343}{619}{2003}.

\bibitem{Centrella01}
J. M. Centrella, C. B. New Kimberly, L. L. Lowe, and J. D. Brown, \AJ{550}{L193}{2001}.

\bibitem{Saijo03}
M. Saijo, T. W. Baumgarte, and S. L. Shapiro, \AJ{595}{352}{2003}.

\bibitem{Ott05}
C. D. Ott, S. Ou, J. E. Tohline, and A. Burrows, \AJ{625}{L119}{2005}.

\bibitem{Ou06}
S. Ou and J. E. Tohline, \AJ{651}{1068}{2006}.

\bibitem{Paschalidis15}
V. Paschalidis, W. E. East, F. Pretorius, and S. L. Shapiro, \PRD{92}{121502(R)}{2015}.

\bibitem{East16}
W. E. East, V. Paschalidis, and F. Pretorius, \CQG{33}{244004}{2016}.

\bibitem{SK04}
M. Shibata and S. Karino, \PRD{70}{084022}{2004}.

\bibitem{SF98} 
N. Stergioulas, and J. L. Friedman, \AJ{492}{301}{1998}.  

\bibitem{OTL04}
S. Ou, J. E. Tohline, and L. Lindblom, \AJ{617}{490}{2004}.

\bibitem{UTea2016}
K. Ury\=u, A. Tsokaros, L. Baiotti, F. Galeazzi, N. Sugiyama, K. Taniguchi, and S. Yoshida, \PRD{94}{101302}{2016}.

\bibitem{DLSS04}
M. D. Duez, Y. T. Liu, S. L. Shapiro, and B. C. Stephens, \PRD{69}{104030}{2004}.

\bibitem{TU07}
A. Tsokaros and K. Ury\=u, \PRD{75}{044026}{2007}.

\bibitem{UT12}
K. Ury\=u and A. Tsokaros, \PRD{85}{064014}{2012}.   

\bibitem{UTG12}
K. Ury\=u, A. Tsokaros, and Philippe Grandcl\'ement, \PRD{86}{104001}{2012}.   

\bibitem{TU13}
A. Tsokaros and K. Ury\=u, \JEM{82}{133}{2013}.  

\bibitem{TU15}
A. Tsokaros and K. Ury\=u, \PRD{91}{104030}{2015}.

\bibitem{TMGRU16}
A. Tsokaros, B. C. Mundim, F. Galeazzi, L. Rezzolla, and K. Ury\=u, \PRD{94}{044049}{2016}.

\bibitem{EFLSTB08}
Z. B. Etienne, J. A. Faber, Y. T. Liu, S. L. Shapiro, K. Taniguchi, and T. W. Baumgarte, \PRD{77}{084002}{2008}.

\bibitem{ELS10}
Z. B. Etienne, Y. T. Liu, and S. L. Shapiro, \PRD{82}{084031}{2010}

\bibitem{EPLS12}
Z. B. Etienne, V. Paschalidis, Y. T. Liu, and S. L. Shapiro,  \PRD{85}{024013}{2012}.

\bibitem{PLES10}
V. Paschalidis, Z. Etienne, Y. T. Liu and S. L. Shapiro \PRD{83}{064002}{2011}

\bibitem{PLES11}
V. Paschalidis, Y. T. Liu, Z. Etienne, and S. L. Shapiro \PRD{84}{104032}{2011}

\bibitem{SN95}
M. Shibata and T. Nakamura, \PRD{52}{5428}{1995}.

\bibitem{BS98}
T. W. Baumgarte and S. L. Shapiro, \PRD{59}{024007}{1999}.

\bibitem{HLL83}
A. Harten, P. Lax, and B. van Leer, \SIAMr{25}{35}{1983}.

\bibitem{L77}
B. van Leer, \JCP{23}{276}{1977}.

\bibitem{CW84}
P. Colella and P. R. Woodward, \JCP{54}{174}{1984}.


\bibitem{EH88}
C. R. Evans and J. F. Hawley, \AJ{332}{659}{1988}.

\bibitem{I1980}
J. Isenberg, \IJMPD{17}{265}{2008}; J. Isenberg and J. Nester, 
in \textit{General Relativity and Gravitation}, edited by A. Held, 
(Plenum, New York 1980), Vol 1.

\bibitem{WM1989}
J. R. Wilson and G. J. Mathews, in Frontiers in Numerical
Relativity, edited by C. R. Evans, L. S. Finn, and D. W. Hobill, 
Cambridge University Press, Cambridge, England, 1989.

\bibitem{BS10}
T. W. Baumgarte and S. L. Shapiro, Numerical Relativity: Solving Einstein’s Equations on the Computer,
Cambridge University Press, 2010.

\bibitem{GS84}
G. Gibbons and J. Stewart, in Classical General Relativity, edited by 
W. B. Bonnor, J. N. Islam, and M. A. H. MacCallum, Cambridge Univ. Press, 1984. 

\bibitem{Beig78}
R. Beig, \PLA{69}{153}{1978}.

\bibitem{GB94}
E. Gourgoulhon and S. Bonazzola, \CQG{11}{443}{1994}.

\bibitem{Abbottetal17}
B. P. Abbott, R. Abbott, T. D. Abbott, \textit{et al.}, \arxiv{1701.07709}

\bibitem{EPHMS}
Z. B. Etienne, V. Paschalidis, R. Haas, P.  M\"{o}sta, and S. L. Shapiro \CQG{32}{175009}{2015}


\bibitem{cactus}
Cactus Computational Toolkit, http://www.cactuscode.org.


\bibitem{Baker06}
J. G. Baker, J. M. Centrella, D. I. Choi, M. Koppitz, and J. R. van Meter, \PRL{96}{111102}{2006}.

\bibitem{Campanelli06}
M. Campanelli, C. O. Lousto, P. Marronetti, and Y. Zlochower, \PRL{96}{111101}{2006}.

\bibitem{NRAR13}
I. Hinder, A. Buonanno, M. Boyle, \textit{et al.}, \CQG{31}{025012}{2013}. 

\bibitem{ABDKPST03}
M. Alcubierre, B. Br\"ugmann, P. Diener, M. Koppitz, D. Pollney, E. Seidel, and R. Takahashi, \PRD{67}{084023}{2003}.


\bibitem{ELSB09}
Z. B. Etienne, Y. T. Liu, S. L. Shapiro, and T. W. Baumgarte, \PRD{79}{044024}{2009}.

\bibitem{FS75}
J. L. Friedman and B. F. Schutz, \AJ{199}{L157}{1975}.

\bibitem{MSB99}
S. M. Morsink, N. Stergioulas,  and S. R. Blattnig, \AJ{510}{854}{1999}.

\bibitem{Miller73}
B. D. Miller, \AJ{181}{497}{1973}.


\bibitem{LOM98}
L. Lindblom, B.J. Owen, and S. M. Mosink, \PRL{80}{4843}{1998}.

\bibitem{LD77}
L. Lindblom and S. L. Detweiler, \AJ{211}{565}{1977}.

\bibitem{Betal05}
L. Baiotti, I. Hawke, P. J. Montero, F. L\"offler, L. Rezzolla, N. Stergioulas, 
J. A. Font, and E. Seidel, \PRD{71}{024035}{2005}.

\bibitem{GR07}
B. Giacomazzo and L. Rezzolla, \CQG{24}{235}{2007}.

\bibitem{DAPRG13}
K. Dionysopoulou, D. Alic, C. Palenzuela, L. Rezzolla, and B. Giacomazzo, \PRD{88}{044020}{2013}.

\bibitem{RRG14}
D. Radice, L. Rezzolla, and F. Galeazzi, \CQG{31}{075012}{2014}.












\bibitem{Shapiro04}
S. L. Shapiro, \AJ{613}{1213}{2004}.

\bibitem{MTW}
C. W. Misner, K. S. Thorne, and J. A. Wheeler, Gravitation, Freeman, New York 1973.

\bibitem{SS03}
M. Shibata and Y. Sekiguchi,  \PRD{68}{104020}{2003}.

\bibitem{FE90}
L.S. Finn and C.R. Evans, \AJ{351}{588}{1990}.










\end{thebibliography}


\end{document}